\documentclass[smallextended]{svjour2}       
\usepackage{graphicx}
\usepackage{alltt}
\usepackage{wasysym}
\usepackage{amssymb}
\begin{document}

\title{Tidal synchronization of close-in satellites and exoplanets. II. Spin dynamics and extension to Mercury and exoplanets host stars.}
\author{Sylvio Ferraz-Mello}
\institute{Instituto de Astronomia 
Geof\'{\i}sica e Ci\^encias Atmosf\'ericas\\Universidade de S\~ao Paulo, Brasil \\
sylvio [at] usp.br}
\titlerunning{Tidal synchronization}

\maketitle

\begin{abstract}
This paper deals with the application of the creep tide theory (Ferraz-Mello, Cel. Mech. Dyn. Astron.  
{\bf 116}, 109, 2013) to the rotation of close-in satellites, Mercury, close-in exoplanets and their host stars. The solutions show different behaviors with two extreme cases: close-in giant gaseous planets, with fast relaxation (low viscosity) and satellites and Earth-like planets, with slow relaxation (high viscosity). 
The rotation of close-in gaseous planets follows the classical Darwinian pattern: it is tidally driven towards a stationary solution which is synchronized with the orbital motion when the orbit is circular, but,  if the orbit is elliptical, it has a frequency larger than the orbital mean-motion. 
The rotation of rocky bodies, however, may be driven to several attractors whose frequencies are $1/2,1,3/2,2,5/2,\cdots$ times the mean-motion. The number of attractors increases with the viscosity of the body and with the orbital eccentricity. The final stationary state depends on the initial conditions. 
The classical example is Mercury, whose rotational period is 2/3 of the orbital period (3/2 attractor). The planet behaves as a molten body with a relaxation that allowed it to cross the 2/1 attractor without being trapped, but not to escape being trapped in the 3/2 one. In that case, the relaxation is estimated to lie in the interval $4.6 < \gamma < 27 \times 10^{-9}\ {\rm s}^{-1}$  (equivalent to a quality factor roughly constrained to the interval $5<Q<50$).
The stars have relaxation similar to the hot Jupiters and their rotation is also driven to the only stationary solution existing in these cases. 
However, solar-type stars may  lose angular momentum due to stellar wind, braking the rotation and displacing the attractor towards larger periods.
Old active host stars with big close-in companions generally have rotational periods larger than the orbital periods of the companions.
The paper also includes the study of the energy dissipation and the evolution of the orbital eccentricity.

\end{abstract}
\def\beq{\begin{equation}}
\def\endeq{\end{equation}}
\def\begdi{\begin{displaymath}}
\def\enddi{\end{displaymath}}
\def\ep{\varepsilon}
\def\epp{\varepsilon^\prime}
\def\eppp{\varepsilon^{\prime\prime}}
\def\App{A^\prime}
\def\Appp{A^{\prime\prime}}
\def\bpp{\beta^\prime}
\def\bppp{\beta^{\prime\prime}}
\def\Cpp{C^\prime}
\def\Cppp{C^{\prime\prime}}
\def\CppQ{C^{\prime 2}}
\def\CpppQ{C^{\prime\prime 2}}
\def\aprmenor{\;\buildrel{<}\over{\sim}\;}    
\def\aprmaior{\;\buildrel\hbox{$>$}\over{\sim}\;}    
\def\defeq{\;\buildrel\hbox{\small def}\over{\,=}\;}    
\def\speq{\hspace{1mm} = \hspace{1mm}}    
\def\FRH{ Ferraz-Mello et al (2008) }
\def\SFM{paper I }
\def\Z{\mathbb{Z}}
\def\hig{}
\def\half{\frac{1}{2}}

\section{Introduction}
The rotation of the celestial bodies has always been one central problem in Astronomy.
The knowledge of the rotation of the celestial bodies is important for a great number of reasons; we may mention the role played by the Earth rotation in the past as time keeper and, currently, the need of accurate ephemerides of planets and satellites to plan astronomical  observation and also to plan the flight of space probes to the environment of some nearby ones. 
Theories of rotational motion involve physical models and parameters, and the comparison of ephemerides to the actually observed motion allows us to better constrain the physics of their interiors.
In addition, the knowledge of their rotational evolution gives clues on the physical conditions in which the studied bodies originated and evolved.

One determinant factor for the rotational evolution of one celestial body is the presence of tidal torques. 
Almost all existing theories of tidal evolution are variants of Darwin's main theory (Darwin, 1880) in which the elastic deformation of the body due to the potential of an external body is delayed because of its viscoelastic nature. In Darwin's theory, he postulated that the lag of each tidal component is small and proportional to its frequency, an assumption founded on a very first approximation of a creep approach that he had developed before, but abandoned (Darwin, 1879). It is worth recalling that following the creep approach, he also multiplied the coefficient of each term by the cosine of the lag, one fact almost forgotten in the variants developed in the second half of the past century. 
The introduction of lags proportional to the frequencies and the neglect of the factor proportional to the cosine of the lag deeply affect the results of the Darwinian theories. For example, it implies that the synchronization cannot be achieved when an elliptical relative orbit is assumed. The final state of the rotation of one body after synchronization is supersynchronous\footnote{Supersynchronous means that the angular rotation velocity is larger than the mean-motion, i.e., the rotation period is smaller than the orbital period. It is often referred to, in the literature, as ``pseudo-synchronous" solution.}: $\Omega\simeq n(1+6e^2)$. 

A simplification introduced by MacDonald (1964), in which the whole body was displaced of a constant lag, was adopted by many authors.
Recent analyses, however, showed that this hypothesis is unphysical and not associated to any reasonable rheology (Williams and Efroimsky, 2012; Ferraz-Mello, 2013b). 

The rheology assumed in Darwin's theory and the unphysical assumptions of the MacDonald's theory were widely accepted almost without criticism. It was only recently that Efroimsky and Lainey (2007) raised the point that the rheology of the Earth as revealed by seismological observations diverges from Darwin's assumption and rather gives support to postulate that, in  planetary satellites and terrestrial planets, lags must depend on the frequency through an inverse power law. 
This new model has been successfully applied to explain the fast despinning of Iapetus (Castillo-Rogez et al. 2011).
For the sake of completeness, it is worth mentioning that more general formulations of Darwin's theory (Kaula, 1964; Ferraz-Mello et al. 2008) exist in which the lags are assumed to occur, but they are kept free of any a priori association to frequencies. 

In this paper, we investigate the predictions of the creep tide theory (Ferraz-Mello, 2012; 2013a; hereafter paper I) for the rotation of celestial bodies in systems with close-in companions.
{The elastic tide is not considered. The forces due to the elastic tide are conservative and torque free. They do not need to be considered because they do not affect the rotation of the bodies and the energy dissipation. Only the anelastic tide needs to be considered.}

The solution given by the creep tide theory shows different behaviors in the two extreme cases: close-in giant gaseous planets, with high relaxation factor (low viscosity) and  {large satellites and terrestrial planets}, with low relaxation factor (high viscosity). 
The rotation of close-in gaseous planets follows the classical Darwinian pattern: it is tidally driven towards a synchronous solution when the orbit is circular, but to a super-synchronous solution (a.k.a. pseudo-synchronous), with an angular velocity ($1+6e^2+\cdots$) times the orbital mean-motion, when the orbit is elliptical\footnote{The tidal torques derived in theories funded on the energy dissipation due to tides instead of lags (e.g. Hut, 1981; Eggleton et al. 1998) are the same as in Darwinian theories with lags proportional to frequencies. They also predict pseudo-synchronous stationary rotation with $\Omega \simeq n(1+6e^2)$.}. 
The rotation of  {satellites and terrestrial planets}, however, may be driven to several attractors whose frequencies are $1/2,1,3/2,2,5/2,\cdots$ times the mean-motion. The number of attractors increases with the viscosity of the body and the orbital eccentricity. The final stationary state depends on the initial conditions and on the eccentricity of the orbits. The well-known case of Mercury, whose rotational period is 2/3 of the orbital period (3/2 attractor), is a consequence of the nonzero orbital eccentricity and of the relaxation factor of the planet (large enough to avoid the 2/1 attractor, but small enough to allow it to be trapped in the 3/2 attractor). {The creep tide alone is able to produce such a result if} the relaxation factor lie in the interval $4.6 - 27 \times 10^{-9}\ {\rm s}^{-1}$  (equivalent quality factor roughly constrained to the interval $5<Q<50$). 
{The fact that Mercury has actually an important azimuthal asymmetry in the mass distribution means that a much smaller dissipation is to be expected (see Makarov, 2012, Noyelles et al. 2014).}

The comparison of the results of Darwinian theories to those of the creep tide theory shows that the approximation adopted by Darwin and by almost all authors in the past century corresponds to large relaxation factors and is not valid for terrestrial bodies, in which case, the relaxation factor may be much smaller than the tidal frequencies (see \SFM). The results of the creep tide theory in the case of terrestrial bodies are rather similar to those obtained by Efroimsky and collaborators (Efroimsky and Williams, 2009; Efroimsky, 2012; Williams and Efroimsky, 2012).

This paper deals with the rotational tidal evolution of close-in planetary satellites, Mercury, close-in  exoplanets and their host stars. 
The stars behave as the hot Jupiters -- they have similar relaxation factors -- and their rotation is similarly driven by tides towards the near synchronous attractor. However, the rotation of solar-type stars is also affected by the loss of angular momentum due to the stellar wind which displaces the attractor towards larger rotation periods; old host stars of solar type with big close-in companions have rotational periods generally larger than the orbital periods of the companions (see Ferraz-Mello et al, 2015)

One point to stress in this introduction is that, as in \SFM, we restrict the theory developed in this paper to the case in which the rotation axis of the two bodies is perpendicular to the plane of their relative motion. This is certainly one restrictive hypothesis. However, several points in the development of the theory (stationary rotations, dissipation, transient motions, ...) are the same in this case and in the general case, and to treat them first in this ``coplanar" case allows us a better understanding of the difficulties that they involve. 

We also stress that no permanent triaxiality of the body is assumed.  By this, we mean that the body  is assumed to respond to external torques as a low Reynolds number fluid. The only existing torques are those resulting from the asymmetries introduced into the body by the tidal forces. In the creep tide theory, the rotation is given by a nonlinear first-order differential equation instead of  the second-order pendulum-like equation of the classical spin-orbit dynamics.  (For another tide theory in which the torques are only a consequence of the elasticity of the body and of the action of the tidal forces, see Bambusi and Haus, 2012).

The paper starts with a recapitulation of the creep tide theory emphasizing the two main 
modifications with respect to the previous version: the explicit consideration of the polar oblateness and a different technique to compute the torques acting on the body (Section 2).  Then, Section 3 presents the equations giving the tidal effects on the rotation of the body, whose stationary solutions are discussed in Section 4. Sections 5 considers the particular stationary solutions whose rotation period is 2/3 of the orbital mean-motion and the cases of Mercury and some extrasolar planets. Section 6 considers the changes in the stationary solutions due to the angular momentum leakage resulting from the magnetic braking of active stars hosting exoplanets. At last, sections 7 and 8 present complementary results on the energy dissipation and circularization of the orbits due to tides.

\section{The creep tide}
The basis of the rheophysical approach (\SFM) is the following: 
We consider a homogeneous body $\tens{m}$ of mass $m$ and assume that, at a given time $t$, the surface of the body is a function $\zeta=\zeta(\widehat\theta,\widehat\varphi,t)$ where $\zeta$ is the distance of the surface points to the center of gravity of the body and $\widehat\theta$, $\widehat\varphi$ are their  co-latitudes and longitudes.  The body is under the action of a tidal potential due to one second body $\tens{M}$ of mass $M$ situated in its neighborhood, in the equatorial plane of $\tens{m}$. The figure of equilibrium of $\tens{m}$ under the action of the tidal potential and rotation may be approximated by a triaxial ellipsoid $\rho=\rho(\widehat\theta,\widehat\varphi,t)$ whose major axis is oriented towards  $\tens{M}$, and whose equatorial prolateness and polar oblateness are 
\begin{equation}\label{eq:prola}
\epsilon_\rho=\frac{a_e-b_e}{R_e}=\frac{15}{4}\Bigg{(}\frac{M}{m}\Bigg{)}\Bigg{(}\frac{R_e}{r}\Bigg{)}^3
\end{equation}
and 
\begin{equation}\label{eq:obla}
\epsilon_z=1-\frac{c_e}{R_e}
\endeq
(see Tisserand, 1891; Chandrasekhar 1969; Folonier et al. 2015).
In the above equations, $a_e,b_e,c_e$ are the ellipsoid semi-axes,  
$R_e$ is the mean equatorial radius of $\tens{m}$ ($R_e=\sqrt{a_e b_e}$),
$r$ is the distance from $\tens{M}$ to $\tens{m}$,
$\overline\epsilon_z$ is the component of the polar oblateness forced by the rotation of $\tens{m}$.
Terms of second order with respect to $\epsilon_\rho$ and $\epsilon_z$ are neglected in the calculations. 

\begin{figure}[t]
\centerline{\hbox{
\includegraphics[height=5cm,clip=]{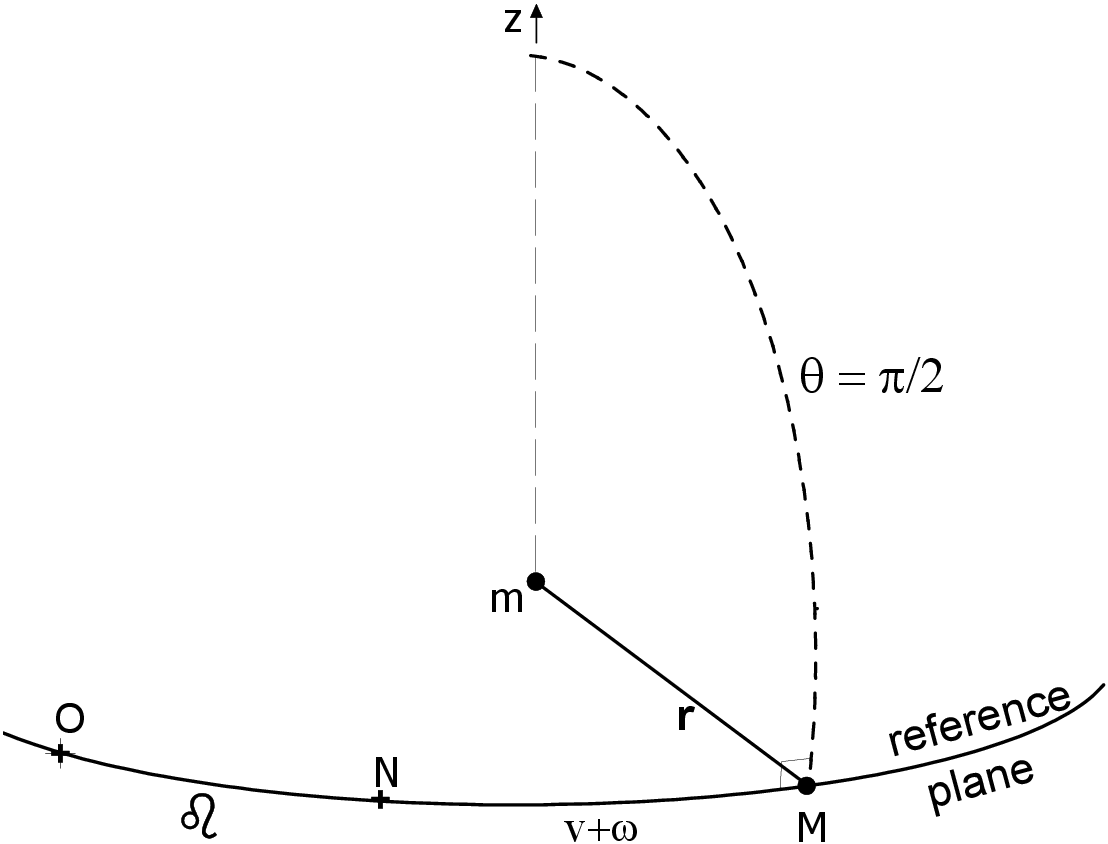}}}
\caption{Spherical coordinates system with origin at the center of $\tens{m}$ and reference plane at its figure equator (see text). $\tens{M}(r,\theta=\pi/2,\varphi)$ is the body creating the tidal potential acting on $\tens{m}$. Angles: $\leo$ is the longitude of the (virtual) node $\tens{N}$; $\omega$ is the argument of the pericenter; $v$ is the true anomaly}\label{fig:angulos}       
\end{figure}
The approach is founded on the Newtonian creeping law
\beq
\dot{\zeta} = -\gamma (\zeta-\rho)
\label{eq:ansatz}\endeq
where the stress in one point of the body was considered as proportional to its (radial) distance to the equilibrium. 
 
The relaxation factor $\gamma$ is a radial deformation rate gradient inversely proportional to the related to the uniform viscosity coefficient $\eta$ through
\begdi   
\gamma=\frac{w \overline{R}}{2\eta} = \frac{3gm}{8\pi \overline{R}^2 \eta},
\enddi
where $w$ and $g$ are the specific weight and gravity acceleration at the surface of the body, and $\overline R$ is its mean radius. 

 It is worth stressing that the creep tide theory is strongly founded on the Newtonian creep, but not restricted to it. As the source of the anelastic component of the tide, the creep is the fundamental part of the theory. It is responsible for the more important tidal effects related to the exchanges of energy and angular momentum (synchronization, dissipation, circularization). It was emphatically treated in \SFM. But the creep tide theory also includes one elastic tide necessary to give the right shape of the body, the precession of the pericenter and the variation of the longitude at the epoch. 

The creep tide theory is virtually identical to the Maxwell model of Correia et al. (2014). The main difference is that in the creep tide theory, the creep tide (or anelastic tide) is calculated separately and then added to the elastic tide, while the Maxwell model of Correia et al. starts with equations that include ab initio the two components of the tide. The only real difference between the two theories is that the creep equation in the Maxwell model adopted by Correia et al. is not given by Eqn.(\ref{eq:ansatz}), but, in the notations of this paper, by $\dot{\zeta} = -\gamma [\zeta-(1-\lambda)\rho]$, where $\lambda$ is one parameter related to the height of the elastic tide. Except for this difference, the two approaches are mathematically equivalent. (For a discussion on the equivalence of the two approaches, see Ferraz-Mello, 2015.)

\subsection{The creep equation}

Equation (\ref{eq:ansatz}) is valid in a reference system co-rotating with the body and such a frame was explicitly used in the first astro-ph preprints of \SFM (Ferraz-Mello, 2012). 
For the actual calculation of the tidal torques, however, it is convenient to note that only relative positions appear in the right-hand side of the creep equation and to use one different frame: the longitude $ \widehat\varphi$ used in the calculations is reckoned from one of the nodes. Hence, $\widehat\varphi$ is a linear function of $t$; in an explicit way,
\beq\label{eq:longitudes}
\widehat\varphi= \widehat\phi_F + (\Omega-\dot{\leo}) t 
\endeq  
where $\widehat\phi_F$ is the longitude of the point in one frame fixed in the body. 
In this equation, $\Omega$ is the angular velocity of rotation of the body, which is assumed to be rotating in the same direction as the relative orbital motion of the two bodies, and $\leo$ is the longitude of the (virtual) node $\tens{N}$ in the non-rotating frame. We keep the denominations of the spatial problem to make easier the generalizations of the present study to 3 dimensions. Obviously, in the context of the coplanar problem,  we could have adopted $\leo\equiv 0$.

Let us consider the creep differential equation corresponding to an equilibrium ellipsoid whose major axis is directed towards $\tens M$: 
\beq\label{eq:edo}
\dot{\zeta}+\gamma \zeta=\gamma\rho = \gamma R_e 
\big(1 + \half \epsilon_\rho \sin^2\widehat\theta \cos (2\widehat\varphi - 2\omega - 2v) - \epsilon_z \cos^2\widehat\theta \big)
\endeq
When $\epsilon_z=0$, Eq. (\ref{eq:edo}) is the same differential equation used in \SFM  with just a small modification coming from the new definition of the angles since, now,  $\widehat\varphi$ was taken with origin in a virtual node $\tens{N}$. 
Therefore, the angle appearing in the arguments in Eq. (\ref{eq:edo}) is $\omega$ instead of $\varpi=\leo+\omega$. 
We note that, in the case of planetary satellites, it is not possible to neglect the motion of the orbit's node and pericenter due, e.g., to the oblateness of the central body. In the integration of Eq. (\ref{eq:edo}) in that case, $\dot{\leo}$ and $\dot\omega$ are considered as nonzero constants.

At this point, we need to stress that, as done in all studies of the effects of tides on rotations, the changing deformation of the body is only taken into account in the calculation of the torques; the reaction of the body to these torques is the same as that  of a rigid body and names as ``figure equator" refer to a surrogate rigid body whose rotation will be affected by the tidal torques. One exception is the theory developed by Williams et al. (2001) that considers the full problem of the rotation of the Moon (not only the tidal perturbations) and where the moment of inertia is decomposed into two parts, one constant -- the ``rigid" moment of inertia -- and the other variable due to tidal and rotational deformations of the body. Here, the influence of the variations of the moment of inertia in the librations in longitude is not considered.

The sequence of the calculations is easy and the same as in Paper I. We first expand the forced terms in the creep differential equation assuming that the functions $r(t)$, $v(t)$ are given by the two-body (Keplerian) approximation. The resulting equation may be written as
\beq\label{eq:edo-Kepl}
\dot{\zeta}+\gamma\zeta 
=\gamma R_e \Big(1+\half\overline\epsilon_\rho \sin^2\widehat\theta 
\sum_{k \in \Z} E_{2,k}\cos \big(2\widehat\varphi +(k-2)\ell - 2\omega\big)
- \overline\epsilon_z \cos^2\widehat\theta 
- \half\overline\epsilon_\rho \cos^2\widehat\theta \sum_{k \in \Z} E_{0,k} \cos k\ell\Big)
\endeq
where 
$E_{q,p}$ are the Cayley functions\footnote{The Cayley functions introduced here correspond to the degree 3 in $a/r$ -- since $\epsilon_\rho \propto (a/r)^3$. More general definitions, corresponding to higher powers will be introduced in Section \ref{sec:dissipation}. 
 {These functions are equivalent to the Hansen coefficients preferred by other authors and the equivalence is given by $E^{(n)}_{q,p}=X^{-n,q}_{2-p}$ (see Correia et al. 2014).}}
(Cayley, 1861; see the Online Supplement)
\beq\label{eq:Cayley}
E_{q,p}(e)=\frac{1}{2\pi}\int_0^{2\pi}\left(\frac{a}{r}\right)^3
\cos\big(qv+(p-q)\ell\big)\ d\ell,
\endeq
\beq
\overline\epsilon_\rho=\frac{15}{4}\Bigg{(}\frac{M}{m}\Bigg{)}\Bigg{(}\frac{R_e}{a}\Bigg{)}^3, 
\endeq
and
\beq
\overline\epsilon_z=\epsilon_z-\half\epsilon_\rho
\endeq
is the polar oblateness due to the rotation of the body (see the Online Supplement).
 
If only the leading terms are kept, we have
\beq
\dot{\zeta}+\gamma\zeta 
= \gamma R_e \big( 1-  \overline{\epsilon_z} \cos^2\widehat\theta \big)
- \frac{1}{4}\gamma R_e \overline{\epsilon_\rho} \cos^2\widehat\theta 
\big(2 +  3  e^2  +   6 e  \cos\ell +  9 e^2  \cos 2\ell \big) +
\endeq\begdi
+\frac{1}{4}\gamma R_e \overline{\epsilon_\rho} \sin^2\widehat\theta \big(
   2         \cos(2 \widehat\varphi - 2 \ell - 2 \omega) 
 +  7 e      \cos(2 \widehat\varphi - 3 \ell - 2 \omega)  
 -  e         \cos(2 \widehat\varphi - \ell - 2 \omega) 
\enddi\begdi
 -  5 e^2  \cos(2 \widehat\varphi - 2 \ell - 2 \omega)  
+  17 e^2  \cos(2 \widehat\varphi - 4 \ell - 2 \omega)\big) + {\cal{O}}(e^3)
\enddi

The above equation corresponds to Eq. (15) of \SFM. The differences with respect to that equation arise from the full consideration of the zonal terms (terms independent of $\widehat\varphi$), which are now  introduced by the polar oblateness $\epsilon_z$ and its dependence on the tidal prolateness $\epsilon_\rho$ (In \SFM only the intersections of the surface with the equatorial plane were taken into account, with some \textit{ad hoc} introduced zonal terms).

Eq. (\ref{eq:edo}) may be written as
\beq	
\label{eq:geral}
\dot{\zeta}+\gamma\zeta = \gamma R_e  + \gamma R_e  \sum_{k\in\Z}\big( {\cal C}_k \sin^2\widehat\theta \cos \Theta_k +  {\cal C}''_k \cos^2\widehat\theta \cos \Theta''_k   \big) 
\endeq
where  ${\cal C}_k$, ${\cal C}''_k$  are constants:
\begin{eqnarray}
{\cal C}_k& \speq & \half \overline\epsilon_\rho E_{2,k} \\
{\cal C}''_k & \speq &-\half \overline\epsilon_\rho E_{0,k} - \delta_{0,k} \overline\epsilon_z 
\end{eqnarray}
($\delta_{0,k}$ is the Kronecker delta), and
$\Theta_k$, $\Theta''_k$ are linear functions of time
\begin{eqnarray}
\Theta_k &\speq & 2\widehat\varphi+(k-2)\ell-2\omega\\
\Theta''_k & \speq & k\ell 
\end{eqnarray}
(N.B. One radial term is missing; see Paper III, Ap. 1)

After the integration, we obtain the forced terms
\beq
\delta\zeta =  R_e \sum_{k\in\Z}\Big( {\cal C}_k  \sin^2\widehat\theta \cos \overline\sigma_k 
\cos (\Theta_k-\overline\sigma_k) + {\cal C}''_k \cos^2\widehat\theta \cos \overline\sigma''_k  \cos (\Theta''_k-\overline\sigma''_k)\Big)
\endeq
where
\begdi
\tan\overline\sigma_k=\frac{\dot\Theta_k}{\gamma} \hspace{1cm} 
\cos\overline\sigma_k=\frac{\gamma}{\sqrt{\dot\Theta_k^2+\gamma^2}}\hspace{1cm}
\sin\overline\sigma_k=\frac{\dot\Theta_k}{\sqrt{\dot\Theta_k^2+\gamma^2}}
\enddi\beq
\tan\overline\sigma''_k=\frac{\dot\Theta''_k}{\gamma} \hspace{1cm} 
\cos\overline\sigma''_k=\frac{\gamma}{\sqrt{\dot\Theta_k^{\prime\prime 2}+\gamma^2}}\hspace{1cm} 
\sin\overline\sigma''_k=\frac{\dot\Theta''_k}{\sqrt {\dot\Theta_k^{\prime\prime 2}+\gamma^2}}
\label{eq:sigmas}
\endeq

As in \SFM, the subtracting constant phases $\overline\sigma_k$ behave as lags, but they are not \textit{ad hoc} plugged constants as in Darwinian theories. They are finite (i.e. not small) exact quantities resulting from the integration of the first-order linear differential equation. We note that being a linear equation, every non-homogeneous term (in the r.h.s.) may be treated separately. It is also worth warning that in the integration, the orbital elements $a,e$, the rotation velocity $\Omega$ and the variations $\dot\omega$, $\dot{\leo}$ are taken as constants. In fact, they are variable. However, their resulting variations are of the order ${\cal O}(\overline\epsilon_z$) and their contributions can be neglected, at least for limited times.

{The free term of the solution, corresponding to the solution of the homogeneous equation 
$\dot\zeta+\gamma\zeta=0$, is transient ($\zeta=Ce^{-\gamma t}$) and will not be considered here. It will be assumed that the past elapsed time is such that these transients were fully damped.} 

The body surface is $R_e+\delta\zeta$ and it is simple to compute the force and torque that it exerts on the external body $\tens{M}$ because $\delta\zeta$ is formed by the bulges of a set of quadrics (which may give positive or negative contributions) superposed to one sphere.
Since these bulges are very thin (they are proportional to $\overline{\epsilon_\rho}$), we may proceed as in Paper I and calculate the attraction of $\tens{M}$ by the resulting composite, as the sum of the forces due to each ellipsoid bulge. The errors of this superposition are of second order w.r.t the oblatenesses (see \SFM).  

In this paper, however, we use a more direct approach.
We substitute the bulges by a thin spherical shell of radius $R_e$ and assume for the mass element at the shell coordinates $(\widehat\theta,\widehat\varphi)$, the sum of the masses of the bulges at that point. The generic mass element in the shell is 
\beq
dm(\widehat\theta,\widehat\varphi)=R_e^2 \mu_{\tens m} \sin\widehat\theta d\widehat\varphi d\widehat\theta \delta\zeta
\endeq
where $\mu_{\tens m}$ is the density of the body. There is a small offset due to the fact that $\delta\zeta$ is the height over a sphere of radius $R$ instead of the mean radius $\overline{R}$, but the changes thus introduced may be neglected. 
The offset is of order ${\cal O}(\overline \epsilon_\rho^2)$.

The contribution of the element $dm$ to the potential in the external point $\tens{P}(r,\varphi,\theta)$ 
is
\beq
dU=-\frac{G dm}{\Delta}
\endeq
where $G$ is the gravitation constant and $\Delta$ is the distance from the element $dm$ to the point $\tens{P}(r,\varphi,\theta)$; 
the potential created by the whole shell is given by
\beq
U=-G R^2\mu_{\tens m} \int_0^{\pi} \sin\widehat\theta d\widehat\theta 
\int_0^{2\pi} \frac {\delta\zeta}{\Delta} d\widehat\varphi
\endeq

The integration is simple and {may be easily computed either numerically or algebraically to the desired precision.} The result is U=$\sum_{k\in\Z}(\delta U_k+\delta U''_k)$, where we have considered separately the contributions of the sectorial {(or, rigorously speaking, tesseral)} and zonal components of $\delta\zeta$ and {neglected terms of higher orders in $R_e/r$}:
\beq
 \delta U_k =-\frac{3Gm R_e^2}{5r^{3}}\ {\cal C}_k\cos\overline\sigma_k
 \sin^2\theta\cos(2\varphi-\beta_k), 
\endeq
\beq
\delta U''_k = -\frac{Gm{R_e}^2} {5r^{3}} {{\cal C}''_k} \cos \overline\sigma''_k (3 \cos^2\theta-1)\cos \beta''_k.
\endeq
The $\beta_k$ are the linear time functions:
\begin{eqnarray}
\beta_k &\speq  &(2-k)\ell + 2\omega +\overline\sigma_k\\
\beta''_k &\speq & k\ell -\overline\sigma''_k.
\end{eqnarray}
We have ommited from $\delta U''_k$ the term $-\frac{Gm} {r} \sum_k{{\cal C}''_k}\cos \overline\sigma''_k \cos\beta''_k$
which is the potential of an inverse square central force proportional to the mass of the $\delta\zeta$-shell. 
This term is obviously an artifact due to the chosen approximations and it must be discarded because the undisturbed central force potential already includes the whole mass of the body. 
It is worth mentioning that such a term only appears when the zonal terms are considered; 
the contribution to the mass of the  terms $\delta\zeta$ with sectorial variation  
 is null because the deficit/excess of mass along adjacent sectors of the shell compensate one another since they are similar but with reversed sign.

It is worth noting that the leading terms in $\delta U_k$ and $\delta U''_k$ are proportional to $1/r^3$. This fact is of importance since it will be responsible for tidal forces inversely proportional to the 4-th power of the distances. 

\subsection{Forces and torques acting on $\tens{M}$} \label{sec:torque}

To obtain the force acting on one mass located at one point, we have to take the negative gradient of the potential at that point and multiply the result by the mass placed on the point. 
Since we are interested in the force acting on $M$ due to the tidal deformation of $\tens{m}$, once the gradient is calculated we can substitute $(r,\theta,\varphi)$ by the coordinates of $\tens{M}$. Hence,

\beq\begin{array}{l@{\speq}l}
F_{1k}&\displaystyle -\frac{9GMm{R_e}^2}{5r^{4}}{\ {\cal C}_k\cos\overline\sigma_k}
\sin^2\theta\cos(2\varphi-\beta_k  ) \vspace{2mm}\\
F_{2k}&\displaystyle \frac{3GMm{R_e}^2}{5r^{4}}{\ {\cal C}_k\cos\overline\sigma_k}
\sin 2\theta   \cos(2\varphi-\beta_k) \vspace{2mm}\\
F_{3k}&\displaystyle -\frac{6GMm{R_e}^2}{5r^{4}}{\ {\cal C}_k\cos\overline\sigma_k}
\sin\theta \sin(2\varphi-\beta_k);\\
\end{array}\endeq
\bigskip

and 

\beq\begin{array}{c@{\speq}l}
F''_{1k}&\displaystyle -\frac{3GMm {R_e}^2} {5r^{4}} {\ {\cal C}''_k} \cos \overline\sigma''_k 
(3 \cos^2\theta-1)\cos \beta''_k \vspace{2mm}\\
F''_{2k}&\displaystyle -\frac{3GMm {R_e}^2} {5r^{4}} {\ {\cal C}''_k} \cos \overline\sigma''_k 
\sin 2\theta\cos \beta''_k \vspace{2mm} \\
F''_{3k}&0. 
\end{array}\endeq
The corresponding torques are
\beq
M_{1k}=0,\qquad   M_{2k}=-rF_{3k}, \qquad   M_{3k}=rF_{2k},
\endeq
that is
\beq\begin{array}{l@{\speq}l}
M_{2k}&\displaystyle \frac{6GMm{R_e}^2}{5r^{3}}{\ {\cal C}_k\cos\overline\sigma_k}
\sin\theta \sin(2\varphi-\beta_k) \vspace{2mm} \\
M_{3k}&\displaystyle \frac{3GMm{R_e}^2}{5r^{3}}{\ {\cal C}_k\cos\overline\sigma_k}
\sin 2\theta   \cos(2\varphi-\beta_k) \\
\end{array}\endeq
\bigskip
and, for the zonal terms, similarly,
\begdi
M''_{1k}=0,\qquad   M''_{2k}=-rF''_{3k}=0, \qquad   
\enddi
and
\beq
M''_{3k}=rF''_{2k}=\displaystyle -\frac{3GMm {R_e}^2} {5r^{3}} {\ {\cal C}''_k} \cos \overline\sigma''_k 
\sin 2\theta\cos \beta''_k .
\endeq

\subsubsection{Forces and torques acting on an equatorial $\tens{M}$}

The variables $\theta$ and $\varphi$  are the co-latitude and longitude of $\tens{M}$ in the frame defined in fig. \ref{fig:angulos}. Since $\tens{M}$ is assumed to lie in the equatorial plane of $\tens{m}$, $\theta=\pi/2$ and $\varphi=v+\omega$.Hence
\beq\begin{array}{l@{\speq}l}
F_{1k}&\displaystyle -\frac{9GMm{R_e}^2}{5r^{4}}{\ {\cal C}_k\cos\overline\sigma_k}
\cos(2v-(2-k)\ell -\overline\sigma_k  ) \vspace{2mm}\\
F_{2k}&0\\
F_{3k}&\displaystyle -\frac{6GMm{R_e}^2}{5r^{4}}{\ {\cal C}_k\cos\overline\sigma_k}
\sin(2v-(2-k)\ell -\overline\sigma_k);\\
\end{array}\endeq
and 
\beq\begin{array}{c@{\speq}l}
F''_{1k}&\displaystyle \frac{3GMm {R_e}^2} {5r^{4}} {\ {\cal C}''_k} \cos \overline\sigma''_k 
\cos (k\ell -\overline\sigma''_k) \vspace{2mm}\\
F''_{2k}& 0 \vspace{2mm} \\
F''_{3k}&0. 
\end{array}\endeq
The corresponding torques are
\beq
M_{1k}=0,\qquad   M_{2k}=-rF_{3k}, \qquad   M_{3k}=rF_{2k},
\endeq
that is
\beq\begin{array}{l@{\speq}l}
M_{2k}&\displaystyle \frac{6GMm{R_e}^2}{5r^{3}}{\ {\cal C}_k\cos\overline\sigma_k}
\sin(2v-(2-k)\ell -\overline\sigma_k) \vspace{2mm} \\
M_{3k}&0
\end{array}\endeq
\bigskip
and, for the zonal terms, similarly,
\beq
M''_{1k}=0,\qquad   M''_{2k}=-rF''_{3k}=0, \qquad  M''_{3k}=rF''_{2k}=0.
\endeq
Again, we have results that differ from those on paper I because of the consideration, here, of the zonal contributions of the actual polar oblateness of the body.


\section{Rotation}\label{sec:rot} 
We use the equation $C\dot{\Omega}=M_2\sin\theta$ (the $z$-component of the torque on $\tens{M}$ is $-M_2\sin\theta$; see \FRH) where $C$ is the moment of inertia with respect to the rotation axis. Hence
\beq\label{eq:omegadot}
\dot{\Omega} = \ \frac{3GM\overline\epsilon_\rho }{2r^{3}}\sum_{k\in\Z} 
E_{2,k}\cos\overline\sigma_k   \sin^2\theta 
\sin\big(2\varphi-(2-k)\ell-2\omega-\overline\sigma_k\big) 
\endeq
where we have simplified the coefficient by using the homogeneous body value $C\simeq \frac{2}{5}mR_e^2$ and introduced the actual values of the Keplerian ${\cal C}_k$ and $\beta_k$.

In the particular case considered where $\tens{M}$ lies in the equator of the deformed body, we have
$\theta=\pi/2$ and $\varphi = v+\omega$ and, after Fourier expansion,
\beq
\dot{\Omega} = \ -\frac{3GM\overline\epsilon_\rho }{2a^{3}}\sum_{k\in\Z}
E_{2,k}\cos\overline\sigma_k   \sum_{j+k\in\Z}
E_{2,k+j} \sin(j\ell+\overline\sigma_k) \label{eq:OmegadotKepl}
\endeq
where the summations are done over all terms of order less than or equal to a chosen $N$. (Remember that $E_{2,k}={\cal O}(e^k)$.)

One important characteristic of this equation, due to the invariance of the torque to rotations of the reference system, is that the right-hand side is independent of the attitude of the body ($\tens{m}$). The arguments of the periodic terms do not include the azimuthal angle fixing the position of the rotating body. Therefore, this is a true first-order differential equation and there are no free oscillations. The corresponding physical librations are forced oscillations. This is totally different from the classical spin-orbit dynamics of rigid bodies where a permanent  azimuthal asymmetry in the mass distribution of the body (potential terms with coefficients $J_{22}$ or $J_{31}$) gives rise to terms including the azimuthal angle in the arguments and the equation to be considered is a second-order differential equation. 

The average  of Eq. (\ref{eq:OmegadotKepl}) with respect to $\ell$ is
\beq\label{eq:OdotKep}
<\dot{\Omega}> = -\ \frac{3GM\overline\epsilon_\rho }{4a^{3}}\sum_{k\in \Z}
E_{2,k}^2   \sin 2 \overline\sigma_k.
\endeq
To truncate at order $N$ (necessarily even), we discard all terms with $|k| > N/2$. 

From eqns. (\ref{eq:sigmas}), we obtain, 
\beq
\cos \overline\sigma_k = \frac{\gamma}{\sqrt{\gamma^2+ (\nu+kn)^2}} \qquad
\sin \overline\sigma_k = \frac{\nu+kn}{\sqrt{\gamma^2+ (\nu+kn)^2}} \qquad
\sin 2 \overline\sigma_k = \frac{2\gamma (\nu+kn)}{\gamma^2+ (\nu+kn)^2}
\endeq
where 
\beq
\nu=2(\Omega-\dot{\leo})-2n-2\dot\omega =2\Omega-2n-2\dot\varpi=2\Omega-2\dot\lambda
\endeq 
is the semi-diurnal frequency\footnote{This definition of $\nu$ is obviously different from that adopted in pure Keplerian approaches (as in paper I) where the possibility of precession of the node and pericenter were not considered, i.e. $\dot\varpi=0$. (N.B. Here, $n=\dot\ell$ is the anomalistic mean-motion.)}. N.B. $\Theta_k=2\Omega+k\ell-2\lambda$ and $\dot\Theta_k=\nu+kn$.

\subsection{Synchronization }\label{sec:spinorb}

Eqn. (\ref{eq:OmegadotKepl}) is the first-order differential equation ruling, in the creep tide theory, the interplay of the orbital motion and the rotation of the body\footnote{We do not use the words `resonance' and `capture' because the dynamics of this approach is not pendulum-like. We rather have attractors and basins of attraction. }. 
It may be written as:
\beq
\dot{\nu} = \ -\frac{3GM\overline\epsilon_\rho }{2a^{3}}\sum_{k\in\Z} 
E_{2,k}   \sum_{j+k\in\Z} 
E_{2,k+j} \left(
\sin 2\overline\sigma_k \cos j\ell + 2\cos^2\overline\sigma_k \sin j\ell\right)
\endeq

For the sake of discussing this equation, we introduce an adimensional variable and a scaled time through 
\beq
y=\frac\nu\gamma; \qquad\qquad x=(\frac n\gamma)(t-t_0) =\frac\ell\gamma. 
\endeq
Hence
\beq
y' = \frac{\dot{\nu}}{n} = -\frac{3GM\overline\epsilon_\rho }{2na^{3}}\sum_{k\in\Z} 
E_{2,k}   \sum_{j+k\in\Z} 
E_{2,k+j} \left(
\sin 2\overline\sigma_k \cos j\gamma x + 2\cos^2\overline\sigma_k \sin j\gamma x \right).
\label{eq:yprime}
\endeq
We also have
\begdi
\cos \overline\sigma_k = \frac{1}{\sqrt{1+ (y+P_k)^2}}
\qquad
\sin  \overline\sigma_k = \frac{(y+P_k)}{\sqrt{1+ (y+P_k)^2}}
\enddi
where
\beq
P_k=\frac{kn}{\gamma}.
\endeq

The new scaled variables allow us to represent solutions corresponding to very different values of the relaxation factor $\gamma$ in only one figure, as in fig. \ref{fig:y}, and compare them. The parameters used to construct these figures are those of a hypothetical Moon in hydrostatic equilibrium ($\overline\epsilon_\rho = 2.8\times 10^{-5})$ 
We note that, when $m \ll M$, the coefficient before the double summation can be reduced to $1.5\ n\overline\epsilon_\rho$ and that it only acts as a scale factor for the derivative. Inside the summations, we only have the eccentricity (through the Cayley functions), the mean anomaly of $\tens{m}$ and the ratio $n/\gamma$ (via $P_k$). Therefore, the results are general enough and depend on the specific problem being considered only through the scaling coefficient.  
The results shown in the figures \ref{fig:y} and \ref{fig:A0A1} are qualitatively valid for any system with same adopted eccentricity (0.0549).    

\begin{figure}[t]
\centerline{\hbox{
\includegraphics[height=4.5cm,clip=]{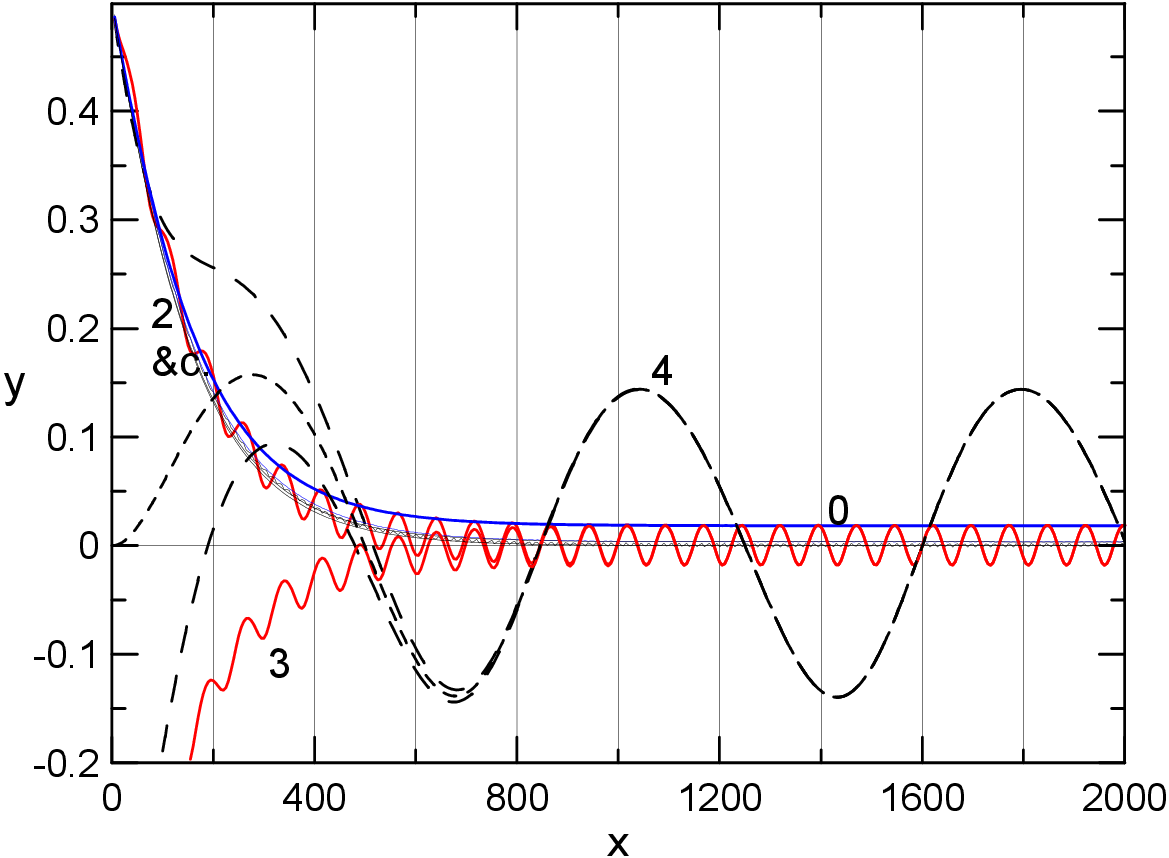}\hspace{3mm}
\includegraphics[height=4.5cm,clip=]{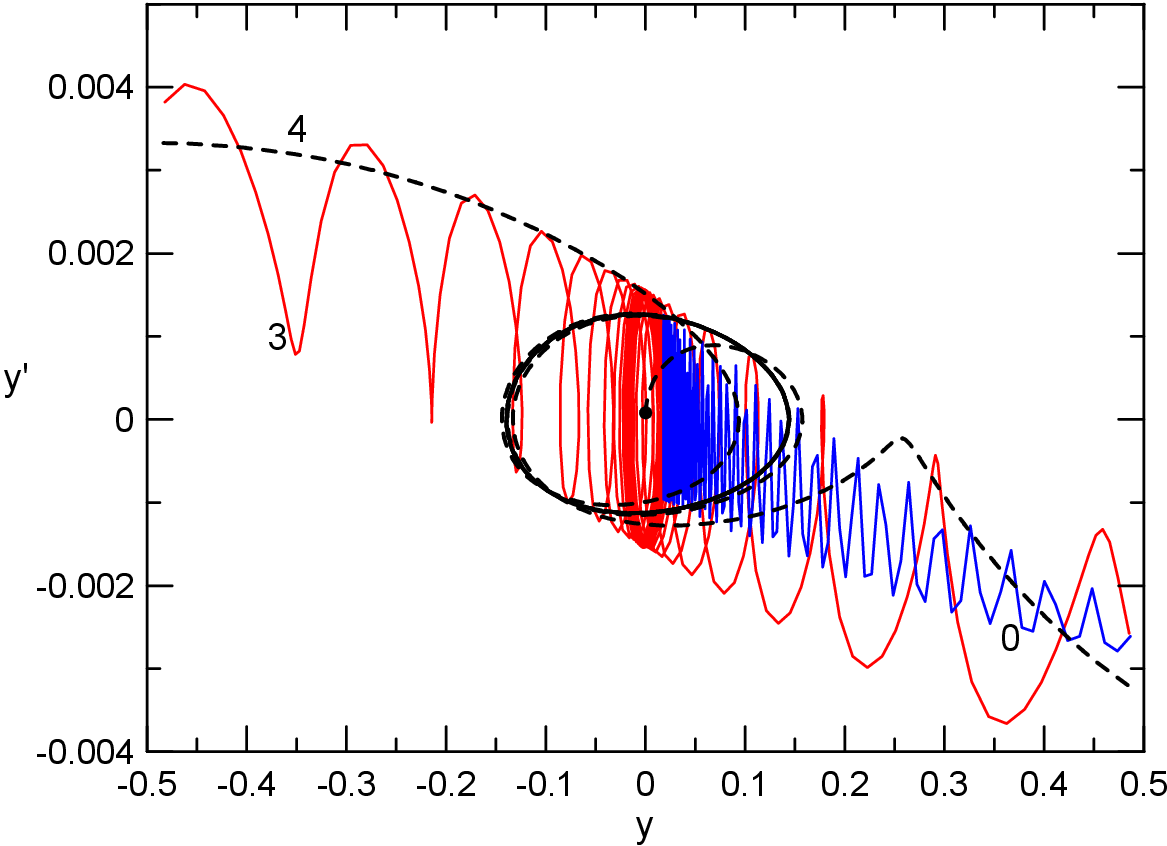}
}}
\caption{\textit{Left}: Solutions of Eqn. (\ref{eq:yprime}) in the neighborhood of $y=0$ for  the initial value $y=0.5$ and several values of $\gamma$ ($\log_{10} n/\gamma=4,3,2,1,0,-1,-2,-3$; see labels);  Solutions with initial values $y=0, y=-0.5$ for $n/\gamma=10^4$ and $y=-0.5$ for $ n/\gamma=10^3$ are also shown. \textit{Right:} Phase portrait of the solutions for $n/\gamma=10^4$, $n/\gamma=10^3$ and $n/\gamma=1$. }     
\label{fig:y}
\end{figure}

\subsection{The neighborhood of the synchronization}
From the above equations, we have $P_0=0$ and
\beq
\cos\overline\sigma_0 = \frac{1}{\sqrt{1+ y^2}} \qquad
\sin\overline\sigma_0 = \frac{y}{\sqrt{1+ y^2}}
\endeq
In the neighborhood of the synchronization, we may assume $\nu <  \gamma $, that is, $ y \ll 1$ and so $\cos \overline\sigma_0 >  {1}/{\sqrt {2}}$ and  $\sin \overline\sigma_0 >  {y}/{\sqrt {2}}$.

\subsubsection{Low-$\gamma$ approximation}\label{sec:lowgamma}

If, besides the above considerations, we have  $\gamma \ll n$ (as in the case of planetary satellites), there follows
\beq
\cos \overline\sigma_{k\ne 0} \simeq \frac{1}{P_k} = {\cal O}(\frac{\gamma}{n}) \ll \cos \overline\sigma_0
\endeq
and the terms with $k\ne 0$ can be neglected.
Eqn. (\ref{eq:yprime}) is then reduced to 
\beq
y' = -\frac{3GM\overline\epsilon_\rho }{2a^{3}n} 
E_{2,0}   \sum_{j\in\Z} 
E_{2,j} \left(
\sin 2\overline\sigma_0 \cos j\gamma x + 2\cos^2\overline\sigma_0 \sin j\gamma x \right)
\endeq
or, taking into account the definitions of $\cos\overline\sigma_0$ and $\sin\overline\sigma_0$,
\beq\label{eq:yprime0}
y' =  -\frac{3GM\overline\epsilon_\rho }{a^{3}n(1+y^2)} 
E_{2,0}   \sum_{j\in\Z} 
E_{2,j} \left(y \cos j\gamma x + \sin j\gamma x \right)
\endeq
or, separating the leading terms,
\begdi
y' =  -\frac{3GM\overline\epsilon_\rho }{a^{3}n(1+y^2)} \Big(
E_{2,0}^2\ y +
E_{2,0}   \sum_{\scriptsize{\begin{array}{c}{j\in\Z}\\{j\ne 0}\end{array}}}
E_{2,j} \left(y \cos j\gamma x + \sin j\gamma x \right)\Big)
\enddi
which is the synchronization differential equation allowing to obtain $y(x)$, that is, $\nu(t)$.

If we neglect the terms with $j\ne 0$, the equation is reduced to
\beq
y' = \frac{\dot{\nu}}{n} = -\frac{3GM\overline\epsilon_\rho }{a^{3}n (1+y^2)}E_{2,0}^2 y 
\label{eq:reduc}\endeq
which is easy to solve. In implicit form, the solution is
\beq
y=y_0 e^{-\half(y^2-y_0^2)} e^{-\frac{\kappa\gamma}{n} (x-x_0)} 
\endeq
or
\beq
\nu=\nu_0 e^{-\frac{1}{2\gamma^2}(\nu^2-\nu_0^2)} e^{-\kappa (t-t_0) }
\endeq
where $y_0=y(x_0)$. It introduces the damping coefficient
\beq
\kappa = \frac{3GM\overline\epsilon_\rho}{\gamma a^{3}}E_{2,0}^2.
\endeq
The solution tends to zero, and the damping time scale ($1/\kappa$) is proportional to $\gamma$. The role of the term $-\half (y^2-y_0^2)$ is to slightly reduce the damping (during the damping, 
$|y|<|y_0|$)

When the terms with $j\ne 0$ are considered, periodic fluctuations which are harmonics of the orbital period of the system are added to the solution (see Fig.\ref{fig:y}). Note that when $n/\gamma=10^{4}$ (dashed lines), the time unit is $10^{-4}$ yr and the period of $\tens{m}$ (the hypothetical Moon) is 754.5 time units, which is the period of the oscillation after the transient phase. When $n/\gamma=10^{-3}$ (red lines), the time unit is 10 times larger and the period of $\tens{m}$ is 75.5 time units.

\subsection{Short-period librations}

The periodic oscillations may be easily obtained from eqn. (\ref{eq:yprime}). We may consider only the term with $k=0$ (since for all others $\cos\overline\sigma_k \ll 1$), introduce $\varphi=v+\omega$,  and transform the true anomaly into mean anomaly. Hence
\beq
y' =  -\frac{3GM\overline\epsilon_\rho }{na^{3}(1+y^2)} 
E_{2,0}   \sum_{j\in\Z} 
E_{2,j} \left(y \cos j\gamma x + \sin j\gamma x \right).
\endeq
or taking into account the values of the Cayley functions and also that $y \ll 1$ and $\overline\sigma_0 \simeq 0$: 
\beq
\dot{\nu}_{|j|=1} \simeq \ \frac{3GM\overline\epsilon_\rho }{a^{3}}\  e \sin\ell 
\endeq

It is worth emphasizing that the amplitude of this forced oscillation is almost independent of $\gamma$ as long as $\nu  \ll \gamma \ll n$.


\subsubsection{High-$\gamma$ approximation}

If $\gamma \gg n$ (as in the case of giant planets and stars), $P_k \ll 1$ and, for $k \ne 0$,  we may use the approximations $\cos \overline\sigma_k \simeq 1$ and $\sin \overline\sigma_k \simeq P_k \ll 1$. 
Therefore, we cannot privilege the terms $k=0$ as done in the low-$\gamma$ approximation, and the synchronization equation then becomes
\beq
y' = \frac{\dot{\nu}}{n} = 
- \frac{3GM\overline\epsilon_\rho}{na^{3}}
  \sum_{\scriptsize{\begin{array}{c}{k\in\Z}\\{k\ne 0}\end{array}}} \sum_{j\in\Z} E_{2,k}   
E_{2,k+j} \sin j\gamma x
\endeq

When the terms with $j\ne 0$ are neglected this equation is reduced to the same equation as before (eqn. \ref{eq:reduc}).

When the terms with $j\ne 0$ are considered, periodic fluctuations which are harmonics of the orbital period of the system are added to the solution. 

\subsection {The stationary solution. Short-period libration. }

Once the transient related to the initial value of $y$ is damped, the solution becomes a periodic function. This is clearly seen in the examples shown where, in the case $n/\gamma= 10^{4}$, 
a large oval around the origin appears (fig. \ref{fig:y} right): the solution is periodic and its average is close to zero. In the other cases, the oval is less apparent because of the squeezing of the adopted time scale. What is also clearly apparent is that in the case $n=\gamma$ (blue curve) the average is shifted to the right. The oval is an attractor. Actually, the projection of a 3-D attracting limit cycle.

We may have recourse to the classical technique of undetermined coefficients to construct the periodic solutions of the given equation. However, we are only interested in the leading terms of the Fourier expansion. Thus, we just assume that $y=A_0+A_1\cos(\gamma x + {\rm phase})$, substitute this approximation in the differential equation, discard the harmonics of second and higher orders and solve to obtain $A_0$ and $A_1$. The results are shown in fig. \ref{fig:A0A1} (left). 
In order to see the actual behavior of the periodic solutions, we show in fig \ref{fig:A0A1} (right) the unscaled value of $\nu$.
This agrees with the non-visibility of the amplitude of the periodic terms in fig.\ref{fig:y} (left) when $\log(n/\gamma) <  2$. 

\begin{figure}[t]
\centerline{\hbox{
\includegraphics[height=4.5cm,clip=]{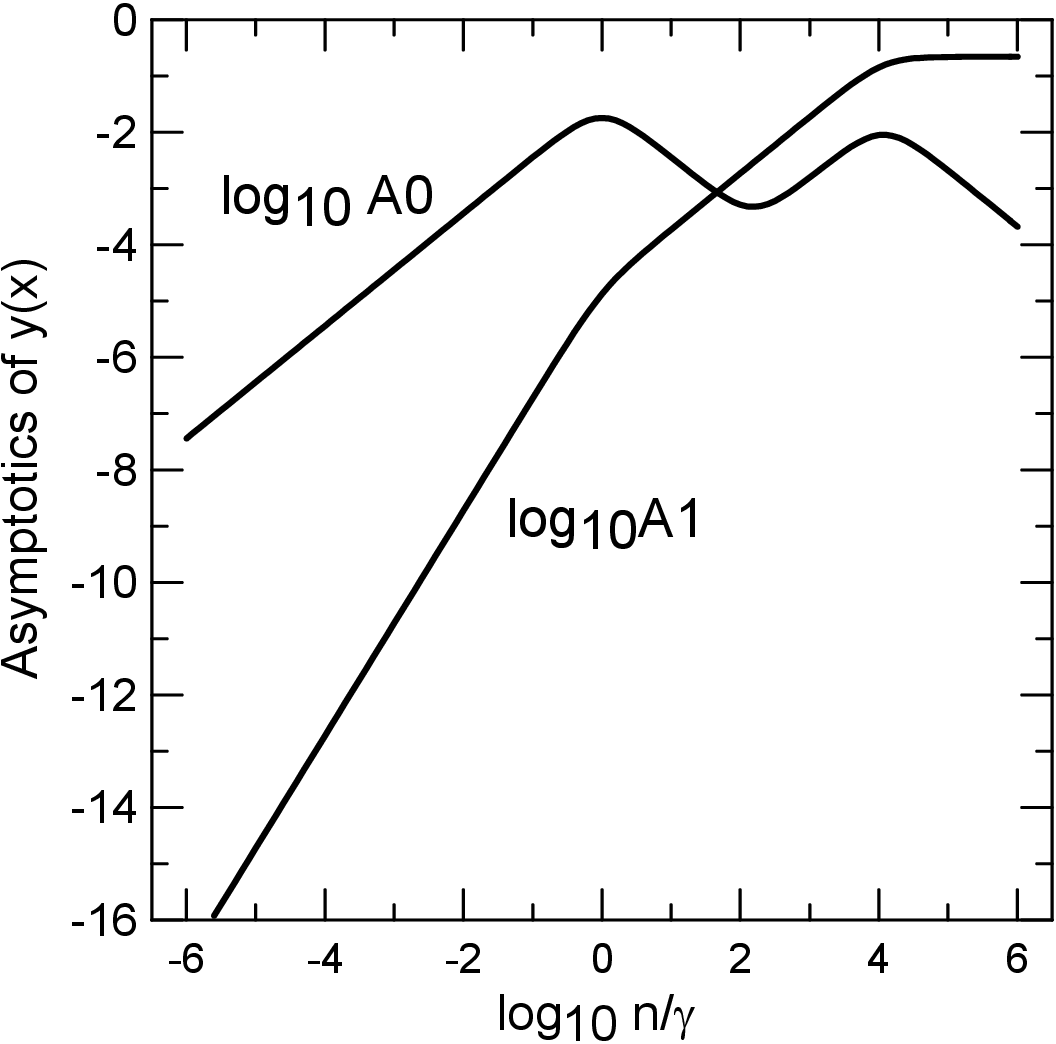}\hspace*{10mm}
\includegraphics[height=4.5cm,clip=]{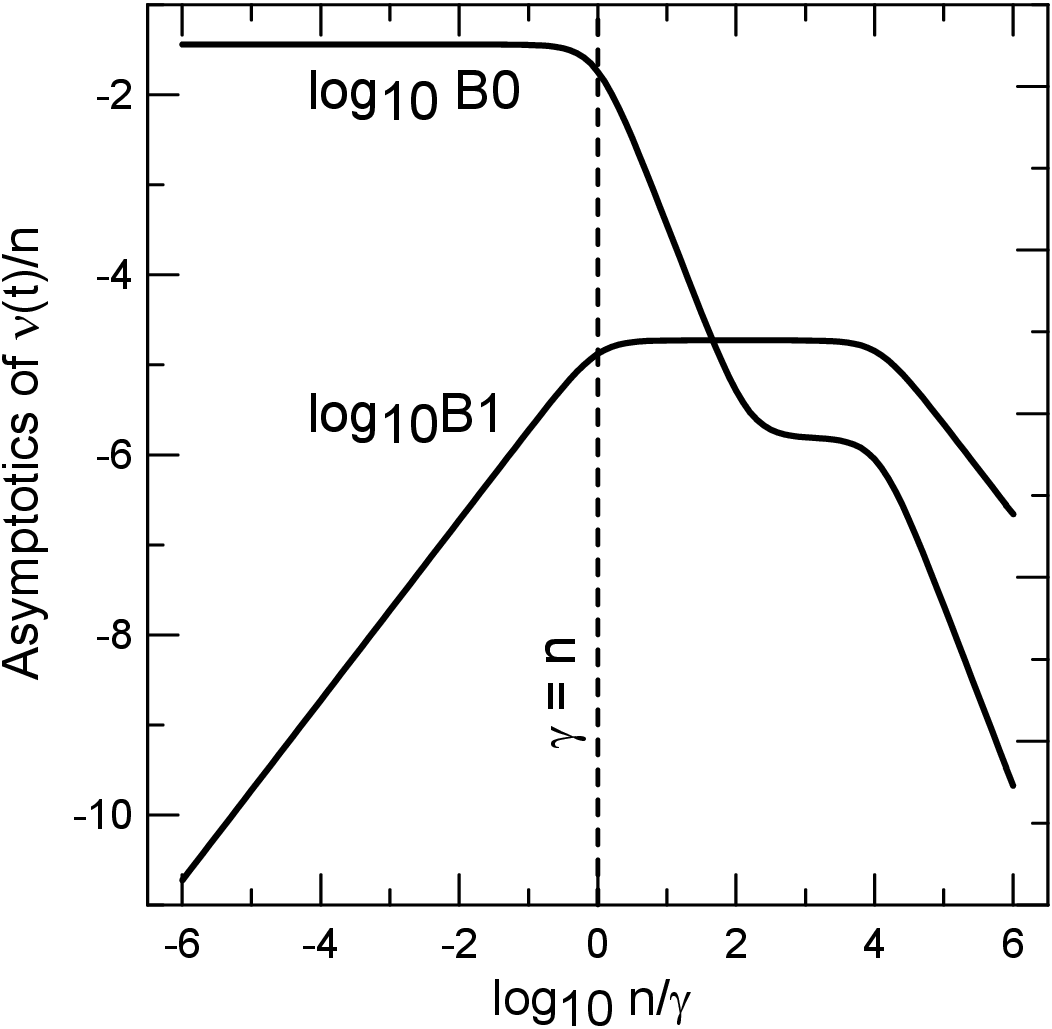}
}}
\caption{\textit{Left}:Mean($A_0$) and amplitude($A_1$) of the asymptotic periodic solution of Eq. \ref{eq:yprime}. \textit{Right:} The same, but unscaled ($B_j=A_j\gamma/n$).}     
\label{fig:A0A1}
\end{figure}

When $n<\gamma$, the asymptotic motion is almost constant and super-synchronous ($\nu>0$, that is $\Omega>n$; the rotation period is smaller than the orbital period) and the offset of the stationary motion ($B_0$) is almost independent of the relaxation factor $\gamma$.
In the opposite case, when $n>\gamma $, Fig. \ref{fig:A0A1}(right) shows that $\nu$ is generally negligible (i.e. $B_j\ll 1) $. 
Besides, when $n \gg \gamma $ the short period oscillation ($B_1$) dominates over the average ($B_0$) what means that the asymptotic motion is dominated by the so-called physical librations. 
This result shows that the use of models limited to the averaged motion for the study of the spin-orbit dynamics of natural satellites and other bodies for which $\gamma \ll n$ is at least very hazardous as these motions are strongly dominated by a periodic oscillation whose period is the orbital period. 

\section{The stationary solutions. Dependence on the eccentricity}\label{sec:stat}

The study of the full set of solutions of this system is impaired by the small values of the derivatives $y'$. In order to get a picture of the solutions space, we construct a map that associates to each value $y$ its increment in one complete period of $\gamma x$ (i.e. one orbital period). Formally, these maps are $y_0=y(\gamma x) \rightarrow y(\gamma x+2\pi)-y_0 $.
The maps are presented using a grid of values of $\nu/n$ ($=y\gamma/n)$ and were computed using an ordinary integrator\footnote{In fact, to obtain the maps a first-order integrator would be enough. $y'$ is too small and we are allowed to assume $y$ constant (that is, $\sigma_k$ constant) in the r.h.s. and just integrate over one cycle of the periodic terms, that is,
\begdi
\Delta(\frac\nu{n})=-\frac{3 \pi M \overline\epsilon_\rho }{(M+m)}\sum_{k\in\Z} 
E_{2,k}^2 \sin 2\overline\sigma_k .
\enddi
The results are almost the same as those shown. This equation is, in fact, just a translation of Eq. (\ref{eq:OdotKep}) to the used adimensional variables.}. They are unidimensional. Because of the small values of the variation of $\nu/n$, the results appear multiplied by $10^6$ in figs. \ref{fig:mapGhigh} and \ref{fig:mapGlow}. 
 
\begin{figure}[h]
\centerline{\hbox{\includegraphics[height=7.0cm,clip=]{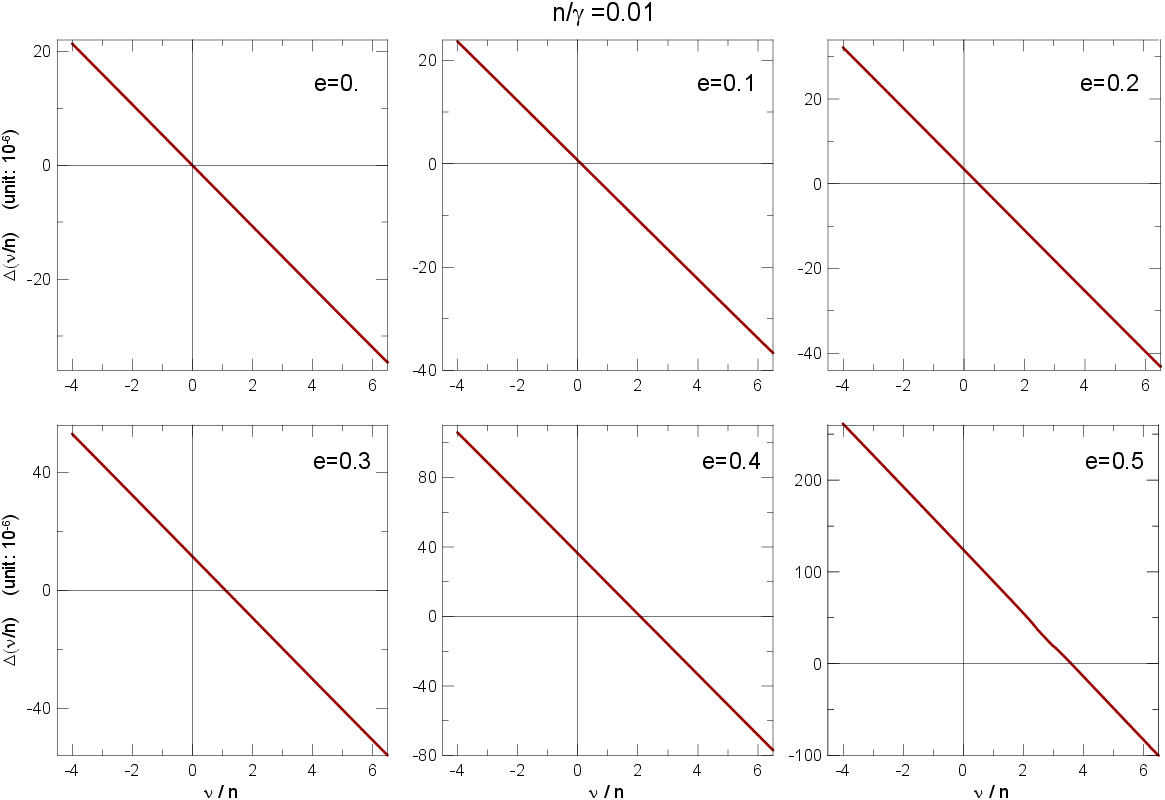}}}\vspace*{5mm}
\centerline{\hbox{\includegraphics[height=7.0cm,clip=]{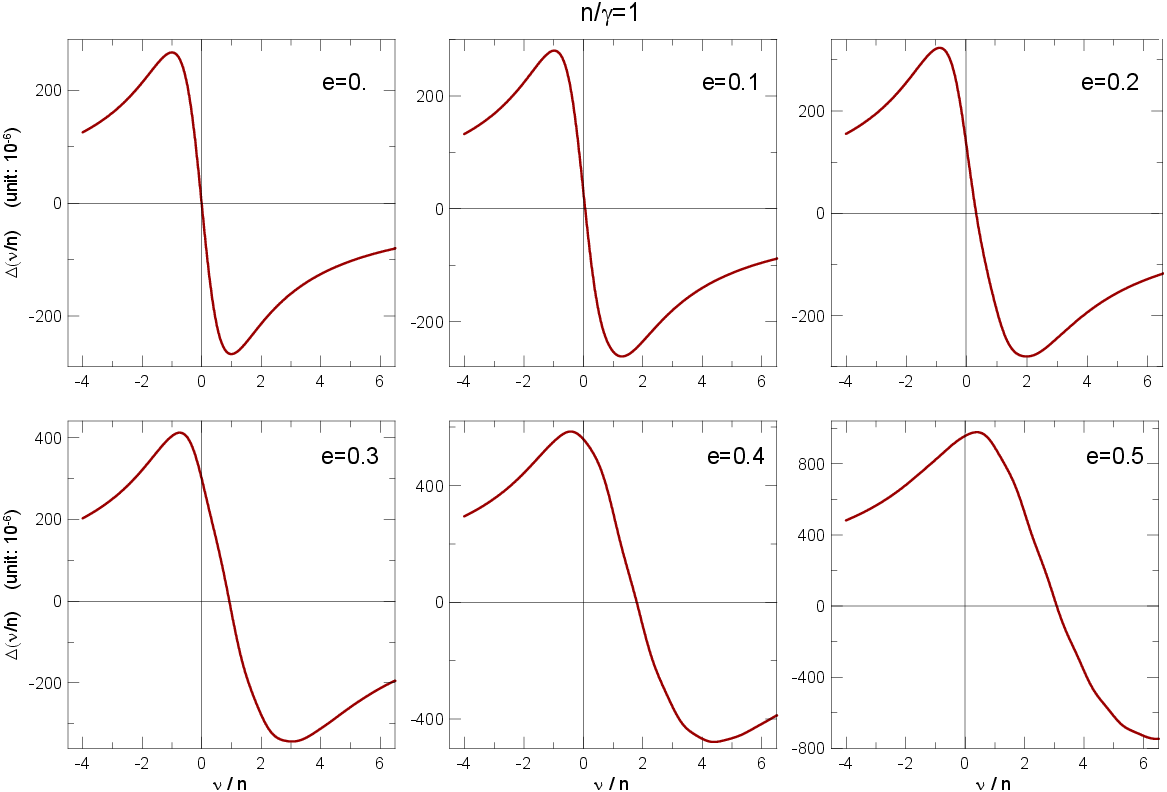}}}
\caption{Maps showing the variation of $\nu/n$ per period for $n/\gamma=0.01$ and $n/\gamma=1$. Remind that for stars and giant gaseous planets, $n/\gamma \ll 1$.}     
\label{fig:mapGhigh}
\end{figure}

\begin{figure}[h]
\centerline{\hbox{\includegraphics[height=7.0cm,clip=]{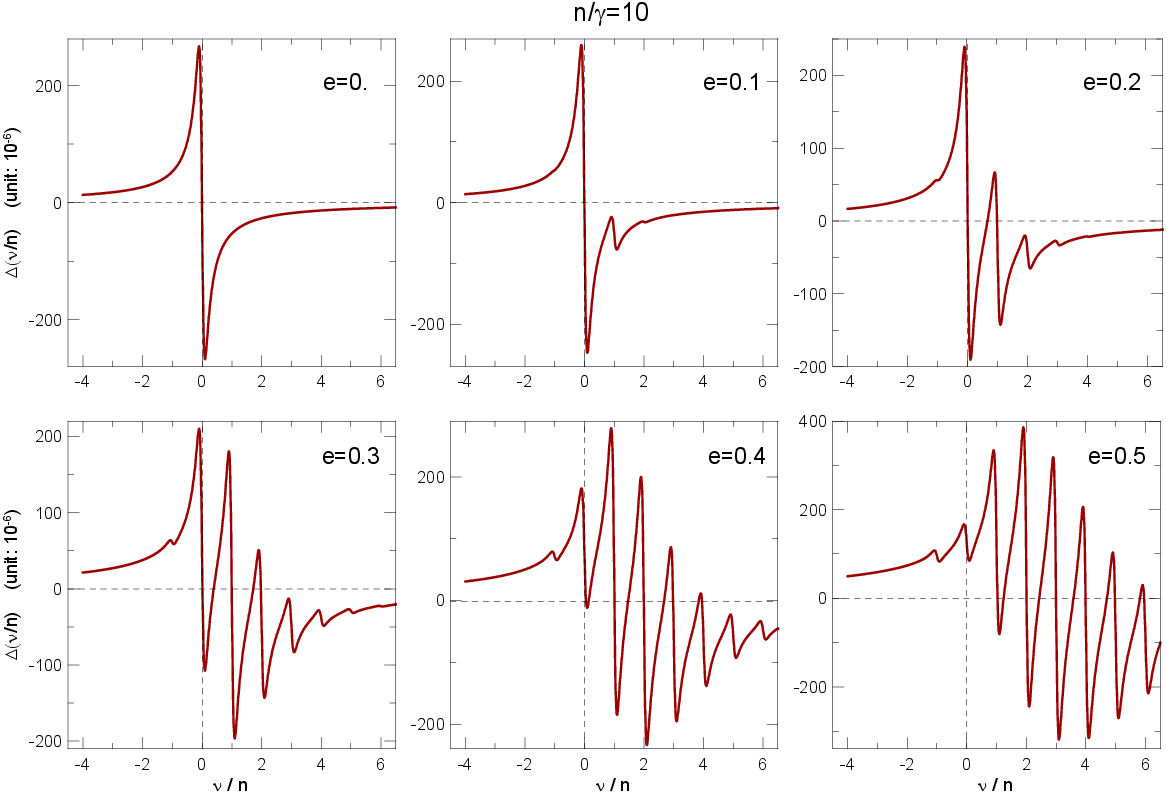}}}\vspace*{5mm}
\centerline{\hbox{\includegraphics[height=7.0cm,clip=]{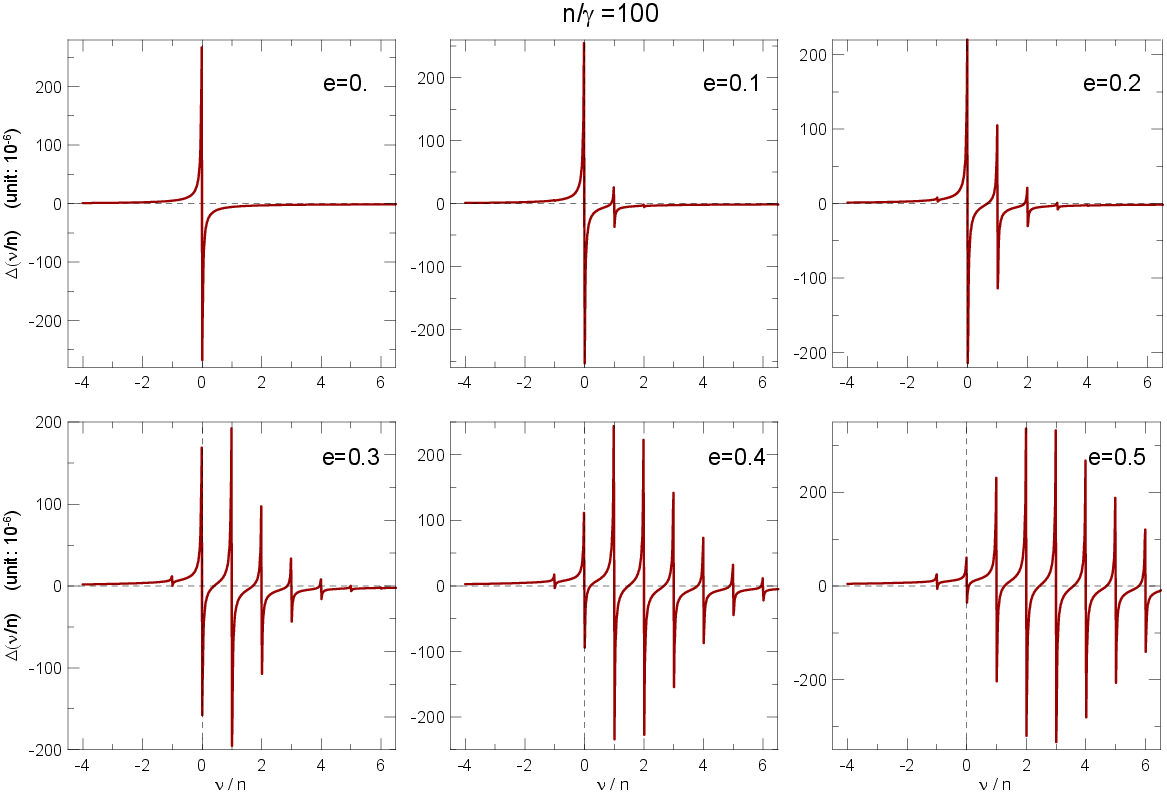}}}
\caption{Maps showing the variation of $\nu/n$ per period for $n/\gamma=10$ and $n/\gamma=100$. Remind that for planetary satellites and terrestrial planets, $n/\gamma \gg 1$}     
\label{fig:mapGlow}
\end{figure}

The maps shown in fig. \ref{fig:mapGhigh} correspond to values of $n$ comparable or less than the relaxation factor $\gamma$. 
The curves intersect the axis $y=0$ (or $\nu/n=0$) just once and show the existence of one and only one attractor (or stationary solution).
As in every other tidal evolution theory, once the damping is over, the motion tends to one attractor solution which, for $e\ne 0$, is supersynchronous
This is what occurs in gaseous giant planets (see \SFM, Table 1). We remember that in this case, the amplitude of the periodic oscillations is negligible (see fig. \ref{fig:A0A1} right). One may see that the intersections are situated at roughly $\nu/n=12e^2$ as given by all Darwin-type theories and by the creep tide theory for the case $n \ll \gamma $ (see \SFM, eqn. 35). For the largest eccentricities, the observed intersection with the axis $\Delta y=0$ is located at a value a little higher than $12e^2$ because of the contribution of the terms of order ${\cal O}(e^4)$. 

The maps shown in fig. \ref{fig:mapGlow} correspond to values of the mean motion $n$ much larger than $\gamma$. The curves intersect the axis $\Delta y=0$ many times and there are many attractors, which are located at $\nu=-n,0,n,2n,\cdots$. However, these attractors are defined by an integral over the period and so are not stationary; once the damping is over the motion is a non-negligible forced libration around the pseudo-stationary values. This is what occurs with  {planetary satellites and Earth-like planets}. Classical examples are the Moon oscillating around $\nu=0$ and Mercury oscillating around $\nu=n$ (i.e. $2\Omega-3n$ oscillating around zero).

Fig. \ref{fig:mapGlow} also shows how the existence of the attractors depends on the orbital eccentricity. When $e=0$, the only attractor is the synchronous solution. When the eccentricity increases, the other attractors at $\nu/n \simeq -1,1,2,3,...$ gradually appear. It is worth noting that the attractor $\nu=0$ does not show, in this case, the offset seen in the cases shown in fig. \ref{fig:mapGhigh}. 
The averages are now very close to the actual synchronization. (The proximity to the actual synchronization is of order ${\cal O}(\gamma/n)^2$.)  

The maps presented in this Section show the same features shown in the plots of the average torque \textit{vs.} rotation frequency 
published by Correia et al. (2014). The similarity is one more consequence of the virtual identity of the creep tide theory and the Maxwell model. 

Makarov and Efroimsky (2013) claim that in the Kaula approach with the introduced lags corresponding to the Efroimsky-Lainey regime, the pseudo-synchronous solution may be unstable. However, in the creep tide theory, the damping $\kappa$ depends on $\gamma$ and $\nu$ through a function which is always positive.  So, the considered stationary solutions are always stable. In the two panels of fig \ref{fig:mapGlow}, $\Delta(\nu/n)$ (i.e. $ <y'>)$  is positive and negative at, resp., the left and right of the attractor. However, looking at fig. \ref{fig:mapGlow}, we may see that between the attractors, there are crossing of the axis $\Delta y=0$ with increasing $\Delta y$. 
These intersections correspond to unstable stationary solutions separating the basins of attraction of different attractors, but these unstable solutions are far from the integer values of $y$. However, Makarov and Efroimsky are correct in the sense that for $\gamma \ll n$ the limit solutions is not an equilibrium solution but a periodic attractor. 

\section{The 3/2 stationary solution} 

The 3/2 solution corresponds to oscillations of the angle $2\Omega-3n$, i.e. $\nu-n$, around zero. We may proceed exactly as in section \ref{sec:lowgamma} since these solutions only exist when $\gamma$ is low, and introduce as new variables: $\nu_1=\nu -n$ and $y_1=y-\frac{n}{\gamma}$. Hence
\begdi
\sin \overline\sigma_k = \frac{(y_1+P_{1k})}{\sqrt{1+ (y_1+P_{1k})^2}}
\qquad
\cos \overline\sigma_k = \frac{1}{\sqrt{1+ (y_1+P_{1k})^2}}
\enddi
where
\beq
P_{1k}=(1+k)\frac{n}{\gamma}.
\endeq
The sequence is exactly the same as in Sec. \ref{sec:lowgamma}, but now, the coefficient 
$\cos\overline\sigma_{-1}$ will be finite while all others will be of the order ${\cal O}(\gamma/n)$.
If the eccentricity is not too small, the leading terms of the equation are
\begdi
y_1' =  -\frac{3GM\overline\epsilon_\rho }{na^{3}(1+y_1^2)} \Big(
E_{2,-1}^2\ y_1 +
E_{2,-1}   \sum_{\scriptsize{\begin{array}{c}{j\in\Z}\\{j\ne 0}\end{array}}}
E_{2,j-1} \left(y_1 \cos j\gamma x + \sin j\gamma x \right)\Big)
\enddi
which is the same equation as the synchronization differential equation. 
The solutions are the same as above, but the damping coefficient now is
\beq
\kappa_1 = \frac{3GM\overline\epsilon_\rho}{\gamma a^{3}}E_{2,-1}^2,
\endeq
which is the order ${\cal O}(e^2)$ since $E_{2,-1}$ is of the order ${\cal O}(e)$ and, for small $e$, much slower than the damping towards the synchronous (or pseudo-synchronous) attractor.


A similar analysis can be done for the 1/2 stationary solution with the difference that, in such case, we will have $E_{2,1}^2$ in the damping. We note that for moderate $e$, $E_{2,1}^2$ is about 50 times smaller than $E_{2,-1}^2$ and this case is thus much less favorable than the 3/2 case. 
For the other stationary solutions, the analyses are similar. However, the powers of the eccentricity entering in the first term become higher. The dampings are much slower and no examples of the other stationary solutions are known in nature. We also remind that the 1/2 solution is shielded by the synchronous attractor. The only way to reach the 1/2 solution is that the body have been in the past in a highly subsynchronous rotation.

\begin{figure}[t]
\centerline{\hbox{\includegraphics[height=7.0cm,clip=]{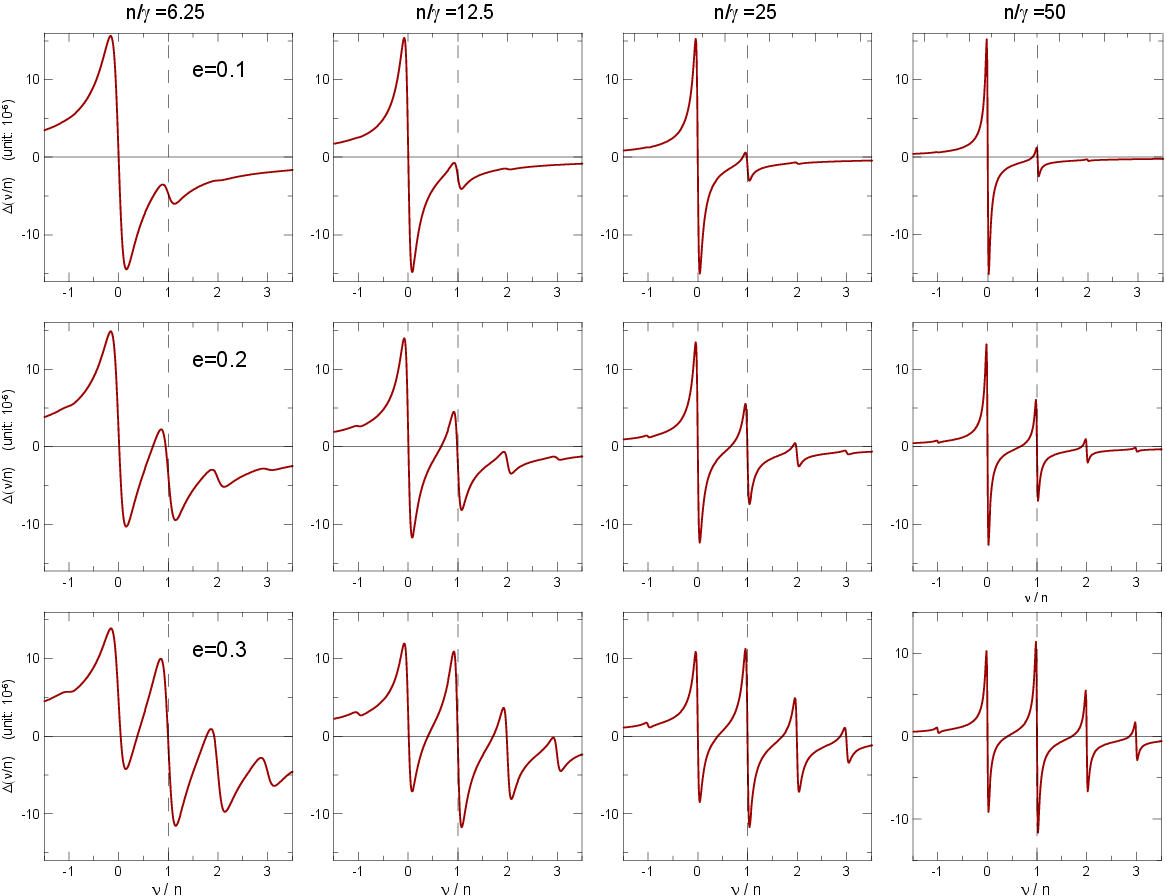}}}\vspace*{5mm}
\centerline{\hbox{\includegraphics[height=7.0cm,clip=]{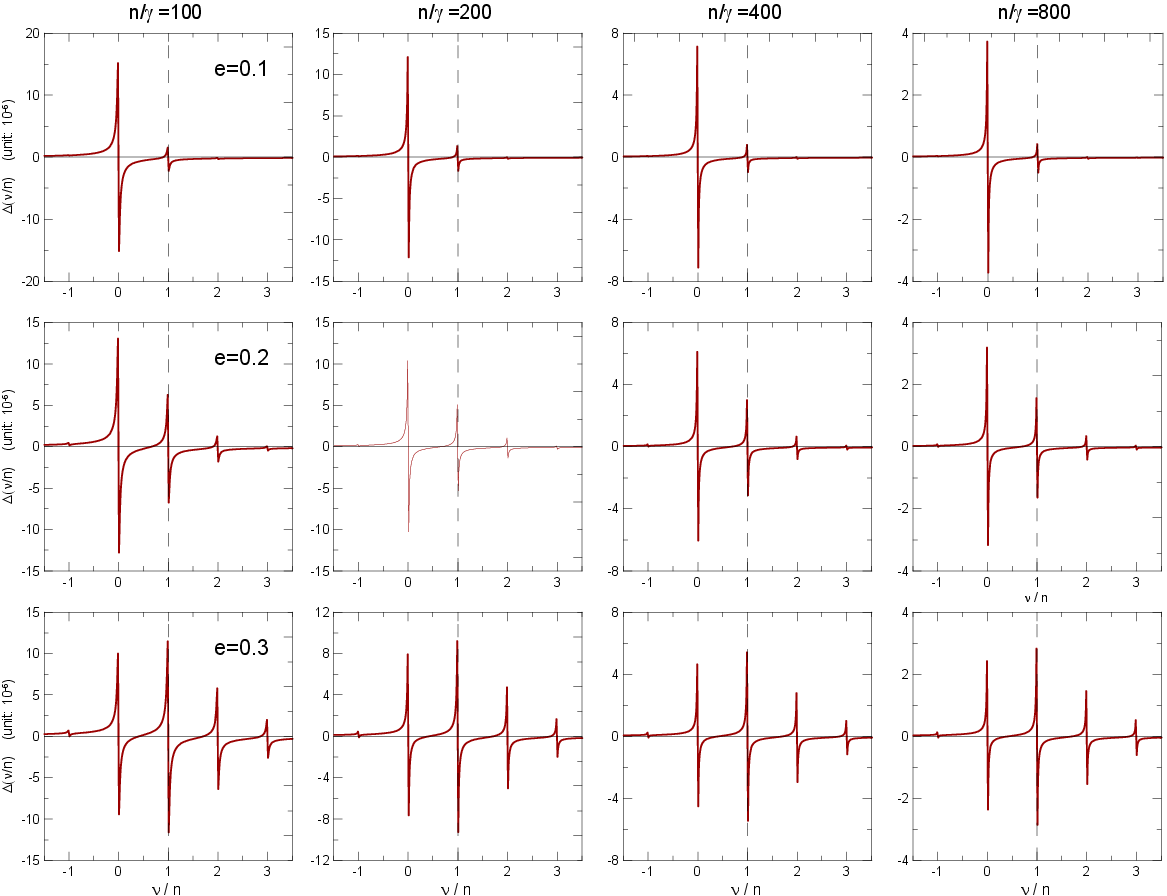}}}
\caption{Maps showing the variation of $\nu/n$ per period for several values of $n/\gamma$ in the case of  {a planet with the same dynamics as Mercury}. The eccentricities were taken at several values to cover the range of secular variation of Mercury's eccentricity.} 
\label{fig:Merc}
\end{figure}

\begin{figure}[t]
\centerline{\hbox{\includegraphics[height=4.5cm,clip=]{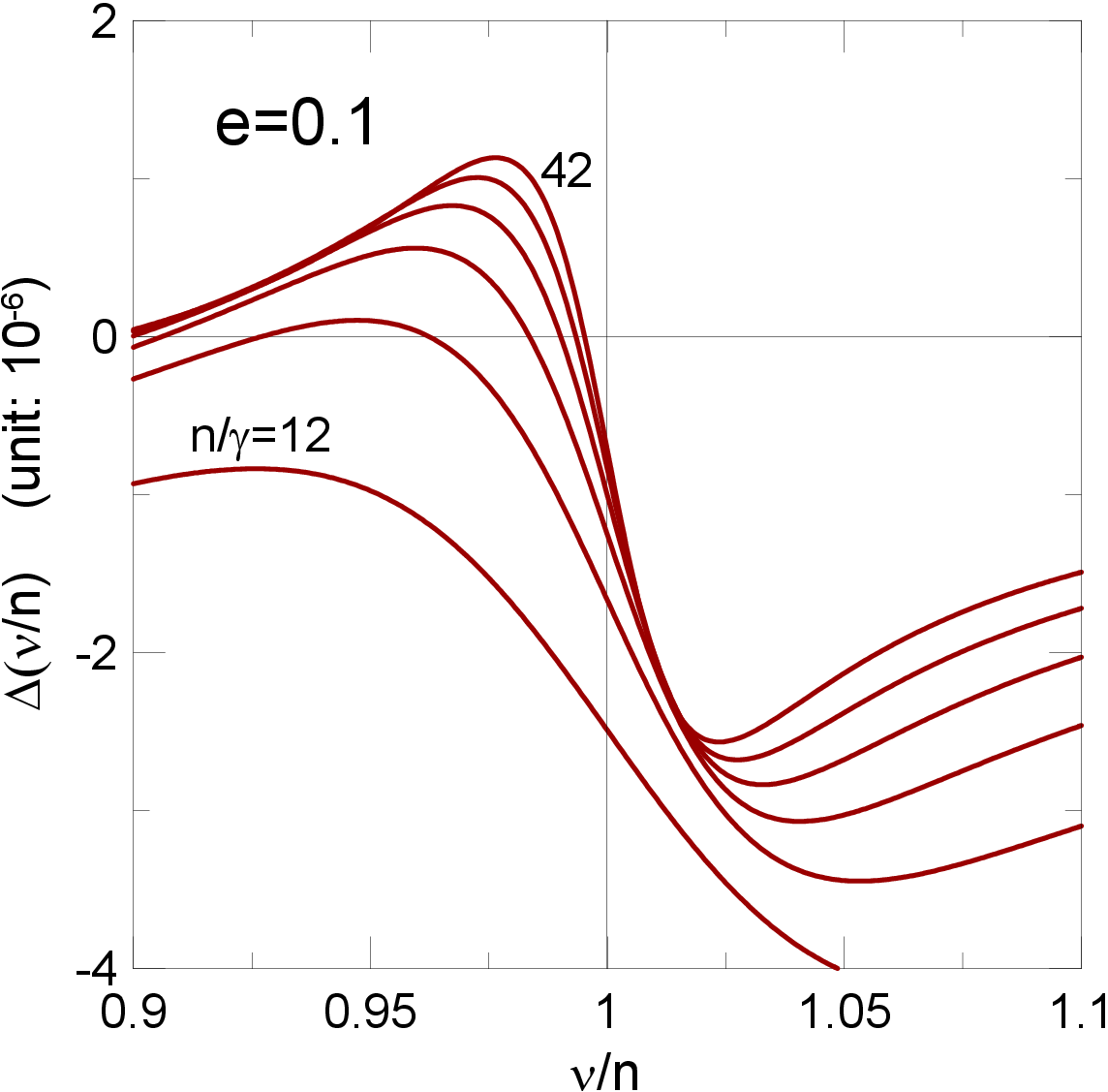}\hspace*{5mm}
\includegraphics[height=4.0cm,clip=]{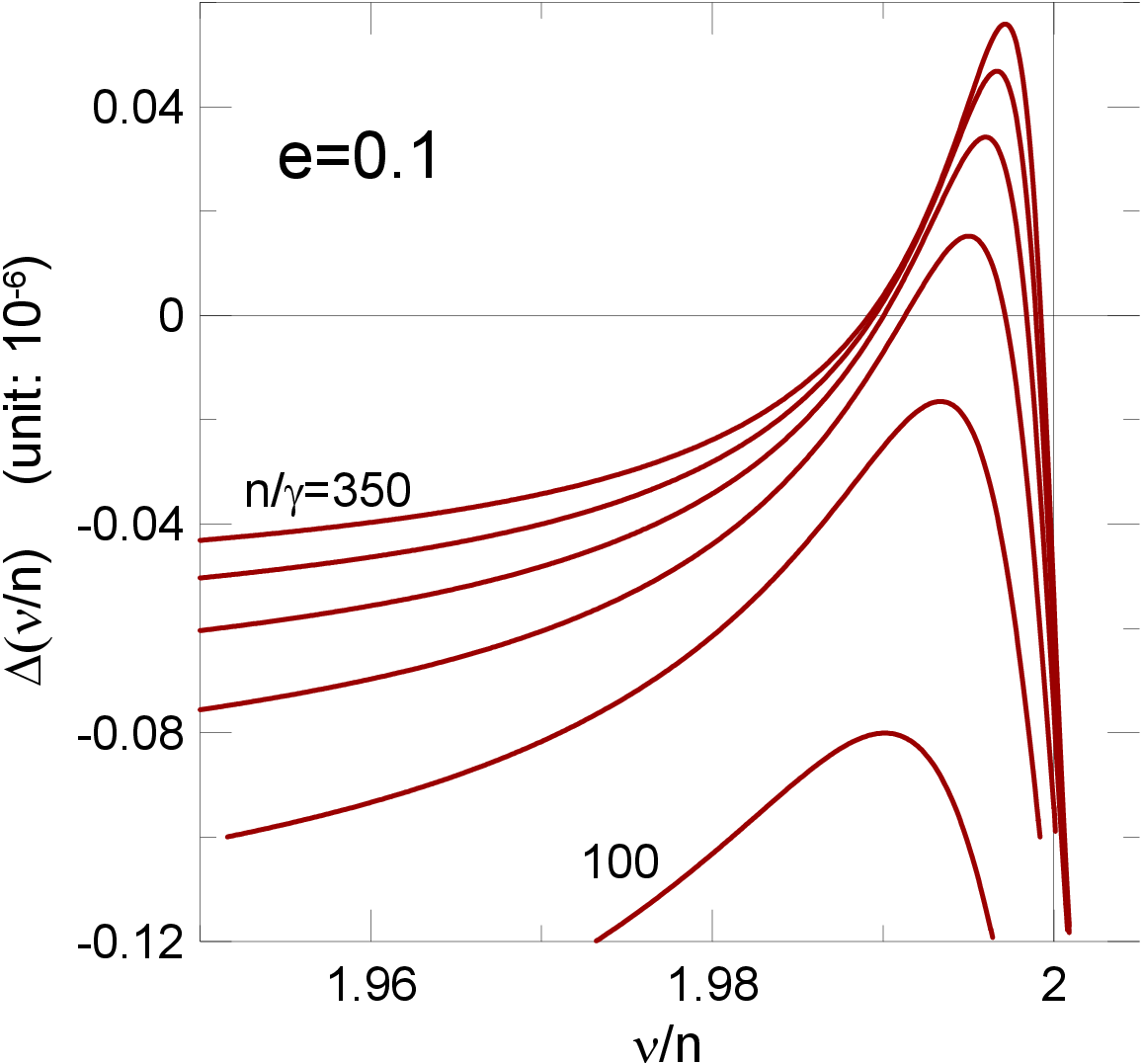}}}
\caption{Barriers in the 3/2 and 2/1 commensurabilities when e=0.1. Variation with $n/\gamma$ increasing in uniform steps from 12 to 42 (left) and from 100 to 350 (right). }     
\label{fig:rise}
\end{figure}

\begin{figure}[t]
\centerline{\hbox{\includegraphics[height=4.5cm,clip=]{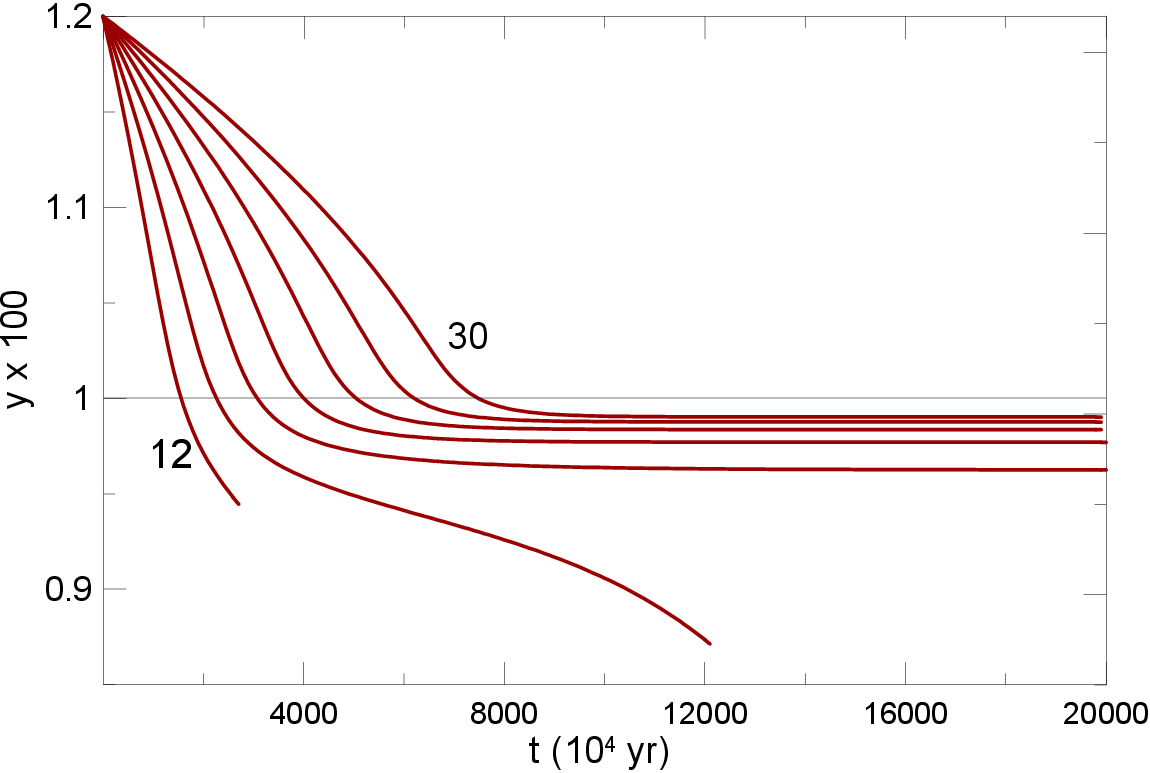}\hspace*{5mm}
\includegraphics[height=4.0cm,clip=]{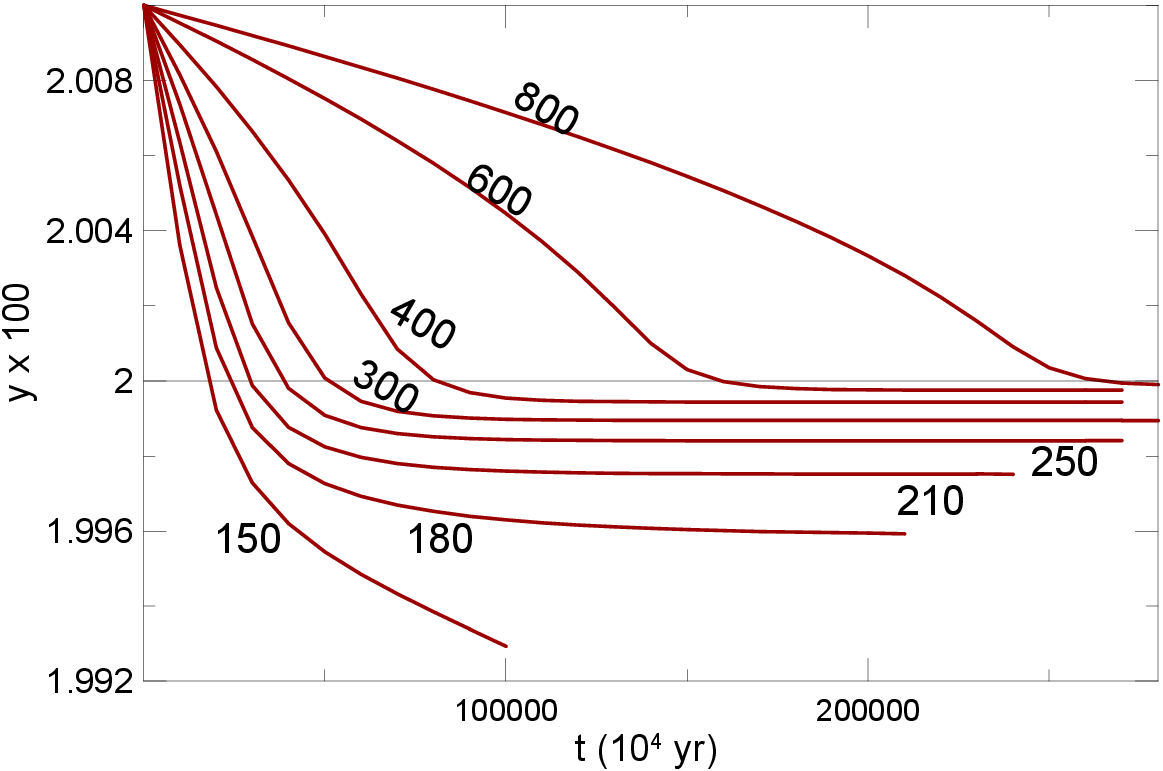}}}
\caption{Simulations with the exact equations in the neighborhood of the 3/2 and 2/1 commensurabilities showing the limiting atractor when e=0.1. In the left $n/\gamma$ increases in uniform steps from 15 to 30. In the right the values of $n/\gamma$ are shown. }     
\label{fig:spvspw}
\end{figure}

\subsection{Application: The rotation of Mercury}\label{sec:Mercury}

Figures \ref{fig:Merc} show the curves $\Delta{y}(y)$ in the case of  one {molten} planet with the same dynamics as Mercury for three values of the orbital eccentricity corresponding to the range of secular variation of Mercury's eccentricity (see Laskar, 1996). The supported scenario is the same of classical theories: Mercury, like the other terrestrial planets, had primordially a rotation much faster than the current one, which slowly evolved due to tidal dissipation up to reach the 3/2 solution where it remained as ``trapped". 
The analysis of the maps for several values of $n/\gamma$, in the neighborhood of the corresponding attractors (see fig. \ref{fig:rise}), and of the results of numerical integration of the exact equations for the same conditions (see fig. \ref{fig:spvspw}) allow us to get an estimation of the relaxation factor of {such emulated} Mercury. 
The fact that it was able to cross the 2/1 solution without being trapped there means that $ n/\gamma < 180$ (i.e. $\gamma > 4.6\times 10^{-9}\ {\rm s}^{-1}$ ). On the other end, the fact that it got trapped into the 3/2 solution, means that $ n/\gamma > 18$ (i.e. $\gamma < 46 \times 10^{-9}\ {\rm s}^{-1}$ ). 
A simple formula to transform these values in the commonly used quality factors $Q$ does not exist and the peculiar rotation of Mercury is different from the rotation states assumed to compute the equivalence formulas given in \SFM.

One may note from fig. \ref{fig:rise} (left) that the 3/2 attractor is not at the exact commensurability. This is more clearly seen in the results of the numerical integrations shown in fig. \ref{fig:spvspw}(left) resulting from simulations in the neighborhood of the 3/2 attractor. The final state of the evolution is below $y=\nu/\gamma=1$. That is, the final solution has a rotation speed slower than $1.5 n$. 
This offset grows in importance for smaller $n/\gamma$. 
It appears small in the maps of fig. \ref{fig:rise} because they were computed for $e=0.1$ but, when the current Mercury's eccentricity (e=0.2056) is adopted, the drifts are yet smaller (see fig. \ref{fig:riseatual})

The fact that no significant drift from the 3/2 commensurability could be measured
(see Margot et al. 2007), indicates a rather higher limit: $ n/\gamma > 30$ (i.e. $\gamma < 27 \times 10^{-9}\ {\rm s}^{-1}$ ).  This puts Mercury's relaxation factor between those adopted for the Moon and Titan (see \SFM) and allows us to constrain Mercury's quality factor $Q$, {in this case}, to the interval 5--50. We may compare this result to the limit  $Q < 100$ determined by 
{Peale and Boss (1977)} as necessary to prevent the trapping in the 2/1 solution.
{ The large Maxwell time (500 yr) adopted by Makarov (2012) and Noyelles et al. (2014) corresponds to values of $\gamma$ much smaller than those found here and is justified by the fact that those authors consider a complete model including the large azimuthal asymmetry in the mass distribution of Mercury, which supply much of the effects attributed to tides when that asymmetry is not taken into account.}

One more comment concerning Mercury's eccentricity in the creep tide scenario is the following: The rotation of Mercury in the 3/2 commensurability is an indication that never in the past, the eccentricity has been much below 0.1. Indeed, for low eccentricities, the 3/2 attractor disappears (see fig.\ref{fig:e05}) or is so weak 
 that with great probability the trapping will not be able to survive the forced periodic variations. Besides, the nature of the phenomenon is such that once the barrier represented by the 3/2 attractor is overcome, the solution tends to the synchronous stationary solution and no longer returns to the neighborhood of the attractor. 

At last, let it be said that the rotation of Mercury is the only observable effect of the tide in Mercury. The variation of $\Omega$ is proportional to $M_\odot (R/a)^3$ and the effect exists because the huge $M_\odot$ compensates the smallness of $(R/a)^3$. The relative variation of the orbital elements: semi-major axis and eccentricity, however, is proportional to 
$M_\odot (R/a)^5$ and the result is strongly dominated by the 5-th power of $(R/a)$ being orders of magnitude smaller than the precision of the nominal values of these quantities appearing in the theories of motion. 

\begin{figure}[t]
\centerline{\hbox{\includegraphics[height=4.0cm,clip=]{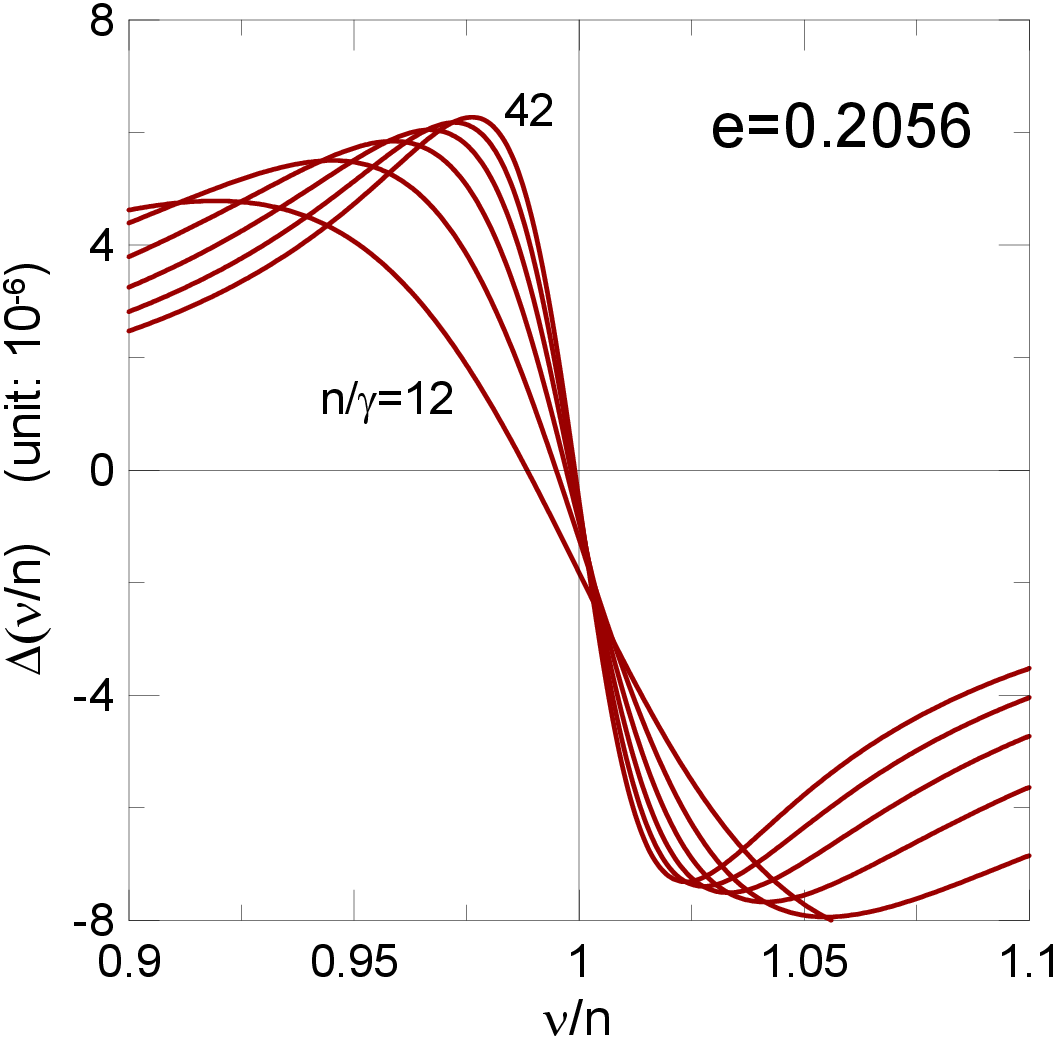}}}
\caption{Barriers in the 3/2 commensurabilities when e=0.2056 (current value). Variation with $n/\gamma$ increasing in uniform steps from 12 to 42}     
\label{fig:riseatual}
\end{figure}

\begin{figure}[t]
\centerline{\hbox{\includegraphics[height=3.5cm,clip=]{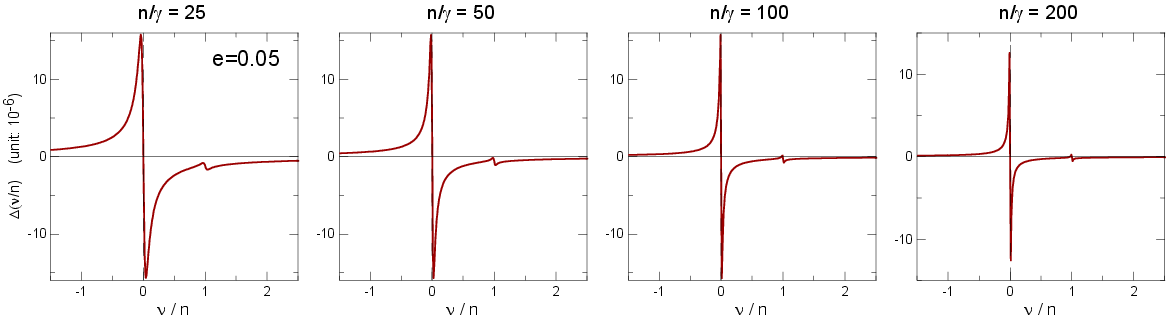}}}
\caption{Maps showing the variation of $\nu/n$ per period for various values of $n/\gamma$ when $e=0.05$}
\label{fig:e05}
\end{figure}

\subsection{3/2 rotation in exoplanets}

The above results may be extended to close-in stiff exoplanets as long as we can assume that they respond to external torques as low Reynolds number fluids. The actual parameters of a given system enter in eqn. (\ref{eq:yprime}) only in the coefficient. The solution behavior discussed in previous section remains the same for other bodies; only the vertical scale changes. Therefore one exoplanet for which $n/\gamma$ is in the interval defined above, $18 < n/\gamma < 180$ in an orbit with eccentricity 0.1, can be trapped in the 3/2 attractor.  
The only additional condition is that it had evolved with almost unchanging orbits and had, at start, a fast rotation. The huge migration ascribed to exoplanet orbits plays against this scenario and the above result must be rather seen from the opposite direction: one planet outside this interval is certainly not in the 3/2 stationary rotation. Fig \ref{fig:rotexo} shows the limits for trapping by the 3/2 attractor of one planet with e=0.1. Below the lines, the 2/1 and 3/2 attractors exist and one body evolving from a fast rotation is trapped by the 2/1 attractor without reaching the 3/2 attractor. Above the lines, the attractors do not exist and the body evolves straight to become synchronous. Between the lines, the 2/1 attractor no longer exists, but the 3/2 attractor exists.

\begin{figure}[h]
\centerline{\hbox{\includegraphics[height=4.0cm,clip=]{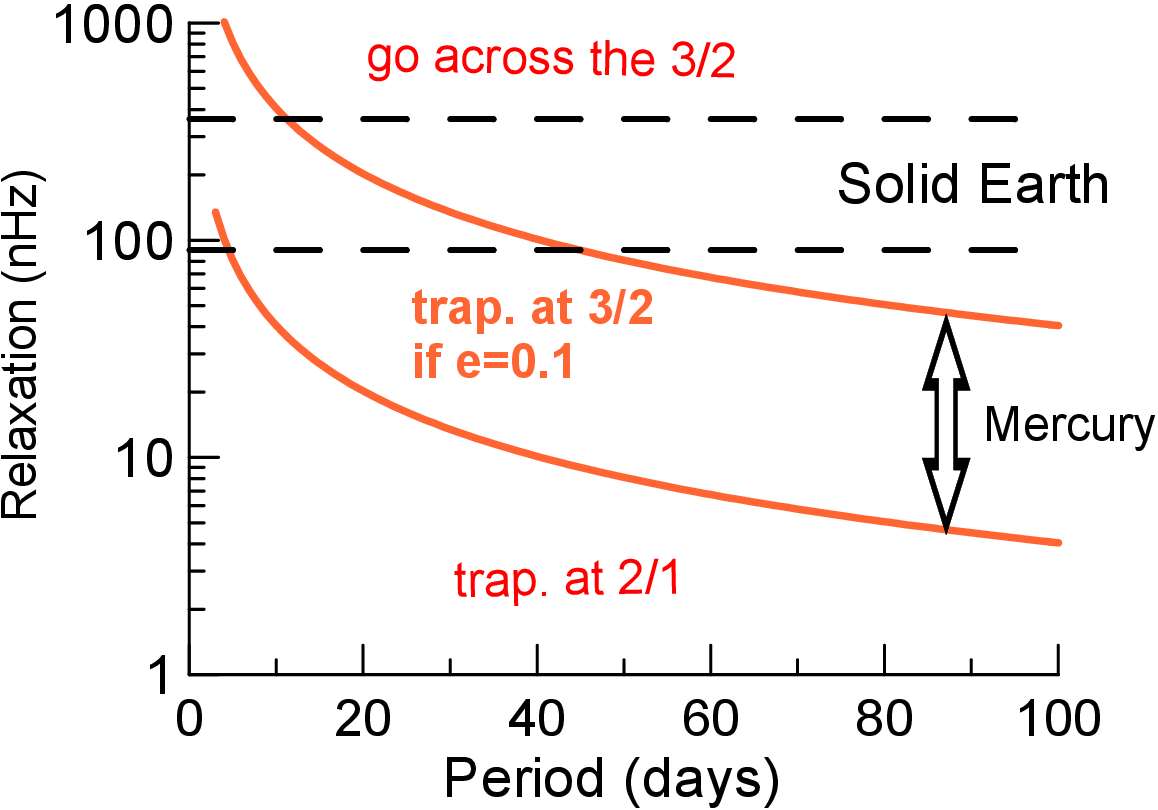}}}
\caption{The red lines show the limits for trapping by the 3/2 attractor of one planet with e=0.1. The relaxation factor of the solid Earth and the limits determined here for Mercury are also indicated. }     
\label{fig:rotexo}
\end{figure}

Makarov et al. (2014), Makarov and Berghea (2014) and Melita (2014) have investigated the possibility of a high-order resonant rotation of the planets GJ 581d and  GJ667Cc whose masses are, respectively, 6.6 and 4.0 Earth masses. 
This possibility can be analyzed in the frame of the creep tide theory. 
One major difficulty, however, is the  fact that the radii of these planets are not known. They can be either super-Earths or mini-Neptunes. 
The relaxation factor of Neptune is in the range $5 - 20\ {\rm s}^{-1}$; {planets similar to Neptune}, but smaller, may have, certainly, $\gamma > 1\ {\rm s}^{-1}$ . This means $n/\gamma \ll 1$ in which case there is only one attractor and the planets may be expected to be in pseudo-synchronous rotation.  

The most interesting case happens if these planets are indeed super-Earths. In such case, the estimated planet masses allow us to assume a solid-Earth-like relaxation factor: $100-500 \times 10^{-9}\ {\rm s}^{-1}$. This may be combined with the orbital periods (66.64 and 28.13 d, resp.) to give $n/\gamma=5-26$ and $n/\gamma>20$. The figures of the previous section show that for such values a 3/2 rotation is indeed possible if the eccentricity is high enough. In this respect, the two planets are different. Due to the secular interaction with GJ667Cd, the eccentricity of GJ667Cc undergoes big periodic variations with the minimum reaching $e \sim 0.05$ (see Makarov and Berghea, op.cit.). In the lowest eccentricity, the 3/2 attractor only exists for $n/\gamma > \enspace \sim\! 100$ (see fig.\ref{fig:e05}). For smaller values, even if the rotation of GJ667Cc was at a given moment trapped by the 3/2-attractor, it would escape when the eccentricity plunge to its minimum value and the rotation would drift towards the synchronous attractor with a negative derivative, without the possibility of returning to the 3/2-attractor. The given condition for a possible 3/2-rotation is equivalent to $\gamma < \enspace \sim\! 25 \times 10^{-9}\ {\rm s}^{-1}$, that is, a relaxation factor of the order of those of Titan or the Moon, which in principle is not expected for a large terrestrial planet. So, if a significant permanent $C_{22}$ does not exist in the planet  potential, the rotation of GJ667Cc may be synchronized with the orbital motion.

In the case of GJ 581 d, the data are more favorable. It has a less variable eccentricity (half amplitude 0.004 \textit{cf.} Makarov et al. 2014) and the best-fit values from the several determinations are similar to present Mercury eccentricity (Hatzes 2013), or larger (Tadeu dos Santos et al. 2012). 
With such eccentricity and $n/\gamma>20$, fig. \ref{fig:Merc} indicates that if this planet is not a mini-Neptune, then we may expect that it be trapped in the 3/2 attractor, or even in the 2/1 one.

The recent discovery of an Earth-sized planet in the habitable zone of Kepler 186 deserves some speculation (Quintana et al. 2014).
In this case, we know the radius of the planet ($1.11\pm 0.4 R_\oplus$) but not its mass. Even if exoplanets often disobey educated  guesses, the expected mass range of Kepler 186 f  ($0.32 < m < 3.77 M_\oplus)$ (Quintana et al. op.cit.) corresponds to non-gaseous bodies. 
In the average hypothesis, the planet could be assimilated to the solid Earth, whose relaxation factor is in the range $90 - 360 \times 10^{-9}\ {\rm s}^{-1}$, of the same order as $n = 560 \times 10^{-9}\ {\rm s}^{-1}$  (the orbital period is 129.9 d). Hence $n/\gamma$ is in the range 1.6 -- 6, too small to allow the existence of the 3/2 attractor unless the planet has a high eccentricity, say 0.1 or larger. If the planet has some liquid parts, high dissipation is likely to occur there (as in the Earth) the equivalent relaxation factor may be smaller and $n/\gamma$ higher, in which case the existence of the 3/2 attractor is possible even if the eccentricity is as small as $\sim 0.05$. The existence of the other non-synchronous attractors depend only on the orbital eccentricity. 
In the extremes of the mass range estimated by Quintana et al (op.cit.), the result will depend critically on the value of $\gamma$. There are no known paradigms with these extreme size and mass and guesses are not possible.

\section{Angular momentum leakage. Host stars}

In solar-type stars hosting planetary systems with large close-in companions, as hot Jupiters or brown dwarfs, the tidal evolution of the system cannot be studied without taking into account a possible loss of angular momentum due to the magnetic wind braking of the star rotation. This angular momentum leakage affects the position of the attractors displacing them toward subsynchronous values. 

One often used model for the stellar wind braking of low-mass stars ($0.5 M_\odot<m<1.1 M_\odot$) is given by the law:
\beq
	\dot\Omega=- f_P B_W\Omega^3
\endeq
where $B_W$ is a factor depending on the star mass and radius through the relation
\beq
	B_W=2.7\times 10^{47} \frac{1}{C}\sqrt{\Big(\frac{R}{R_\odot} \frac{M_\odot}{m} \Big)}\qquad \qquad   ({\rm cgs \enspace units})
\label{eq:Bouvier}
\endeq
(see Bouvier et al. 1997). The factor $f_P$ was introduced by P\"atzold et al.(2012) in the study of the planet of the sub-giant CoRoT-21 to take into account that the braking given by Bouvier's law may be excessive in that case. This may be the case even for some stars similar to the Sun  {and it was used in the study of several other CoRoT systems by Carone (2012).}
The estimates of the loss of angular momentum of the Sun are in the interval $ 3 - 6  \times 10^{30}$ g cm$^2$ s$^{-2}$ while the above value of the numerical coefficient used in Eq. (\ref{eq:Bouvier}) corresponds to $6.6  \times 10^{30}$ g cm$^2$ s$^{-2}$. For low-mass stars, the value can also be much smaller. For these stars, the adopted coefficients in Eq. (\ref{eq:Bouvier}) are in the range $1.2 \times 10^{45} - 1.1 \times 10^{47}$ g cm$^2$ s (see Irwin et al. 2011). The justification and the details of the Bouvier's formula are out of the scope of this paper. (For a comprehensive review, see Bouvier, 2013). For the application concerned here, it is enough to know that the above form of the law is valid after the star has completed its contraction (the moment of inertia $C$ no longer changes), is fully decoupled from the disk and no significant mass loss needs to be considered. Besides, the rotation period of the star may be larger than some saturation value (roughly 1.8 days).

If we adopt the normalized variables introduced in Sec. \ref{sec:spinorb}, the braking equation becomes
\beq
y^\prime = -\frac{2 f_P B_W}{n} (\frac{\gamma y}{2}+n )^3.	
\endeq

\begin{figure}[t]
\centerline{\hbox{\includegraphics[height=3.0cm,clip=]{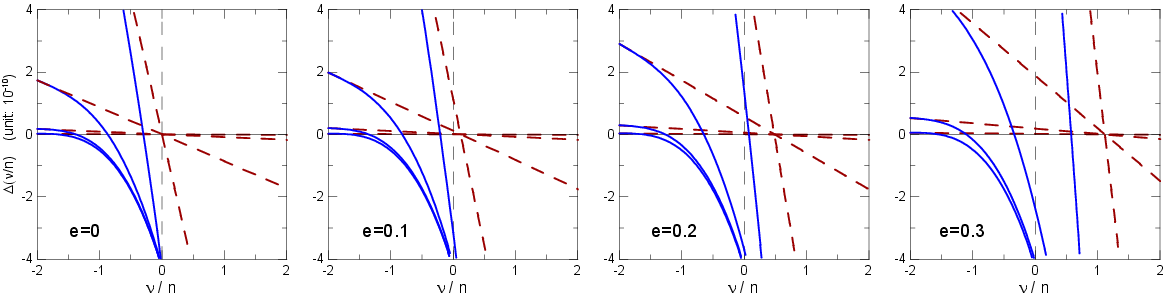}}}
\caption{Location of the pseudo-synchronous solution when $n/\gamma =10^{-6}, 10^{-5}, 10^{-4}, 10^{-3}$, respectively (the curves corresponding to the smaller $n/\gamma$ are those for which the inclination at the intersection with the axis $\Delta=0$ is smaller.
The solutions corresponding to intersections at the left of the origin are sub-synchronous.  
Brown dashed lines: Without braking; Blue solid lines: With braking ($f_P=1.0$)} 
\label{fig:mapa-15x}
\end{figure}

When the braking is added to the synchronization equation of Sec. \ref{sec:spinorb}, the intersections of the map showing the variation of $\nu/n$ per period exoplanets are strongly displaced to the left, that is, to negative values of $\nu/n$. 
The stationary solutions become sub-synchronous. 
Figure \ref{fig:mapa-15x} shows the solutions for values of $n/\gamma$ typical of host stars, for several values of the eccentricity. The maps constructed without the braking are shown in the background (dashed lines) to stress the big changes produced by the angular momentum leakage. The study of some hypothetical cases with a Sun-like star hosting hot Jupiters shows that the rotation of the star is strongly affected in all cases with an initial $a<0.04$ AU, i.e. initial period less than 2 days (Ferraz-Mello et al. 2015).

\section{Energy variation and dissipation}\label{sec:dissipation}
To complete this analysis, it is worth studying the energy variation and the dissipation of energy in the rotating bodies.  
As in our previous papers on tidal theories, we consider only the bulk dissipation that can be predicted by a mere application of the energy conservation principle. The internal mechanisms responsible for it (see  {Ogilvie and Lin 2004, 2007}, Efroimsky and Makarov, 2014,  {Makarov and Efroimsky, 2014} and references therein) are not considered here.  {We have also not yet considered the case of differentiated bodies, in which some parts may be much more efficient to dissipate energy than others (see Remus et al. 2015)}. 

The time rate of the work done by the tidal forces acting on $\tens{M}$ is
\beq
\dot{W}=\mathbf{F}\mathbf{v}= \sum_{k\in \Z} \Big(
(F_{1k}+F''_{1k})\frac{nae\sin v}{\sqrt{1-e^2}}+F_{3k}\frac{na^2\sqrt{1-e^2}}{r}\Big),
\endeq
that is
\beq\begin{array}{ll}
\dot{W}=&-\frac{9GMmnae R_e^2}{10r^4\sqrt{1-e^2}}\sum_{k\in \Z} {\ \cal C}_k\cos\overline\sigma_k
\Big(\sin\big(3v-(2-k)\ell - \overline\sigma_k\big) - \sin\big(v-(2-k)\ell - \overline\sigma_k\big) \Big)\vspace{2mm}\\
&-\frac{6GMmna^2 R_e^2\sqrt{1-e^2}}{5r^5} \sum_{k\in \Z}{\ \cal C}_k\cos\overline\sigma_k
\sin\big(2v-(2-k)\ell - \overline\sigma_k\big)\vspace{2mm}\\
&+\frac{3GMmnae R_e^2} {10r^4\sqrt{1-e^2}} \sum_{k\in \Z} {\ \cal C}''_k \cos \overline\sigma''_k 
 \Big( \sin \big(v + k\ell -  \overline\sigma''_k\big) +  \sin \big(v - k\ell +  \overline\sigma''_k\big)  \Big).
\end{array}\endeq
(Remember that, in the planar case, $\phi=v+\omega$ and $\theta=\pi/2$). 
Taking into account the auxiliary formulas presented in the online Appendix B.3, the above equation is equivalent to
\begdi
\dot{W}=\frac{GMmn R_e^2}{5a^3} \sum_{k\in \Z}\sum_{j\in \Z}
\Big(3 {\cal C}_k\cos\overline\sigma_k (2-k-j) E_{2,k+j} \sin (j\ell+\overline\sigma_k)  
\enddi\beq
+{\cal C}''_k \cos \overline\sigma''_k 
(k+j)E_{0,k+j} \sin( j\ell+ \overline\sigma''_k) \Big),
\endeq
the time average of which is
\beq
<\dot{W}>=\frac{GMmn R_e^2}{10a^3} \sum_{k\in \Z} \Big( 3 {\cal C}_k (2-k) E_{2,k} \sin 2\overline\sigma_k +{\cal C}''_k kE_{0,k} \sin 2\overline\sigma''_k \Big)
\endeq
or, considering the 
definitions of ${\cal C}_k$ and ${\cal C}''_k$,
\beq\label{eq:dotW}
<\dot{W}>=\frac{GMmn R_e^2\overline\epsilon_\rho}{20a^3}  \sum_{k\in \Z}\Big(3 (2-k) E_{2,k}^2 \sin 2\overline\sigma_k - kE_{0,k}^2 \sin 2\overline\sigma''_k  \Big).
\endeq

It is worth noting that the first part of the above expressions (coming from the sectorial terms) are the same found in Paper I. The second part arises from the full consideration of the zonal terms of $\zeta$ in the present approach. It is worth noting that the zonal terms appear, in the above equation,  multiplied by $k$ and so, the term coming from ${\cal C}''_0$ does not contribute to the average energy variation. As expected, the zonal term of the energy variation depends only on the variation of the polar oblateness due to the tidal deformation; it does not depend on $\overline\epsilon_z$, rotational component of the polar oblateness.

If we introduce 
\begdi
\sin 2\overline\sigma_k=\frac{2\gamma (\nu+kn)}{\gamma^2+(\nu+kn)^2},\qquad
\sin 2\overline\sigma''_k=\frac{2\gamma kn}{\gamma^2+k^2 n^2},
\enddi
into Eq. (\ref{eq:dotW}), it becomes
\beq\label{eq:dotWps}
<\dot{W}>=\frac{GMmn R_e^2\overline\epsilon_\rho}{10a^3}  
\sum_{k\in\Z}
\left(3 (2-k) E_{2,k}^2 \frac{\gamma (\nu+kn)}{\gamma^2+(\nu+kn)^2}
 - kE_{0,k}^2 \frac{\gamma kn}{\gamma^2+k^2n^2} \right).
\endeq

\subsection{Synchronous and pseudo-synchronous rotation}
In the particular case of a synchronous rotation, $\nu=0$ and the above equation becomes
\beq\label{dotWsync}
<\dot{W}_{\rm sync}>=\frac{GMmn R_e^2\overline\epsilon_\rho}{10a^3}  
\sum_{\scriptsize{\begin{array}{c}{k\in\Z}\\{k\ne 0}\end{array}}}
\frac{\gamma kn}{\gamma^2+k^2n^2}\left(3 (2-k) E_{2,k}^2  - kE_{0,k}^2  \right).
\endeq
where the terms with $k=0$ were excluded from the summation just to emphasize the fact that their contribution is null, and that, in these cases, the work rate is of the order ${\cal{O}}(e^2)$.

In the synchronous and pseudosynchronous stationary solutions, $<\dot{W}>$ is negative. 
Indeed, the sign of $<\dot{W}_{\rm sync}>$ is determined by the dominating term coming from $k=-1$ in the above sum (note that $E_{2,-1}^2 \gg E_{2,1}^2$). The term $k=-1$ is also dominating in Eqn. (\ref{eq:dotWps}) if $\nu$ is small enough. However, in such case, the contribution of the term coming from $k=0$ cannot be discarded. It is positive, but smaller than the contribution coming from $k=-1$ and not able to change the sign of the result. In the low-$\gamma$ approximation ($\gamma \ll n$), the leading term of the resulting series is
\begdi
<\dot{W}_{\rm ps,low}>\simeq -\frac{33 GMm\gamma R_e^2\overline\epsilon_\rho}{10 a^3} e^2 <0.
\enddi
This formula gives a good approximation in the case of low $\gamma$ (e.g. for satellites) and very small eccentricities, but we note that the complete formula may be easily used to compute $<\dot{W}>$ with a good precision in general cases, and must be preferred.

\subsubsection{Dissipation}

As discussed above, in the synchronous and pseudosynchronous stationary solutions, $<\dot{W}>$ is negative. 
This means that, in these cases, the orbital motion  is being braked, i.e. the system is loosing orbital energy.
In the study of the tides raised in $\tens{m}$, the body $\tens{M}$ may be considered as a material point and, therefore, the whole mechanical energy lost by the system may be converted into heat inside $\tens{m}$\footnote{To obtain the dissipation in the other body, the equations are the same, but with the meanings of $\tens{M}$ and $\tens{m}$ interchanged}.

In the case of a stationary rotation, $\dot\Omega \simeq 0$ and therefore the energy variation associated with the rotation of $\tens{m}$ is almost zero. If we neglect the small variation of $\Omega$ due to the tidal evolution of the semi-major axis, the orbital $\dot{W} (<0)$  represents the whole variation of the mechanical energy of the system and thus, its modulus gives the rate of energy dissipation.

For instance, in the case of Io, the estimated heat flux through Io's surface is at least 2.5 W/m$^2$ (Spohn, 1997) corresponding to a total heat dissipation of 100 TW. To get this flux using Eq. (\ref{eq:dotWps}), we adopt $\gamma=550 \times 10^{-9}\ {\rm s}^{-1}$, which is slightly larger than the value estimated in Paper I but yet inside the error bar of that estimation ($490 \pm 100 \times 10^{-9}\ {\rm s}^{-1}$) and thus in agreement with the value of Io's acceleration determined by Lainey et al. (2009). 

The values of the dissipation for eccentricities from 0.1 to 0.3 in the pseudosynchronous stationary solution and in the synchronous solution are shown in Fig.\ref{fig:Wps}. One may note that the dissipation in the synchronous solutions is higher than the dissipation in the stationary pseudo-synchronous solution of same eccentricity,  {as already shown in the frame of the classical theories by \FRH and Levrard (2008).
Just for memory, we remind that these authors have also shown} that when forced terms due to a permanent equatorial asymmetry of the body are added, the energy variation in the resulting synchronous stationary solution is approximately the same as that of the pseudosynchronous stationary rotation. However, permanent equatorial asymmetries are not expected to occur in hot Jupiters or in planet-hosting stars.

\begin{figure}[t]
\centerline{\hbox{\includegraphics[height=5.5cm,clip=]{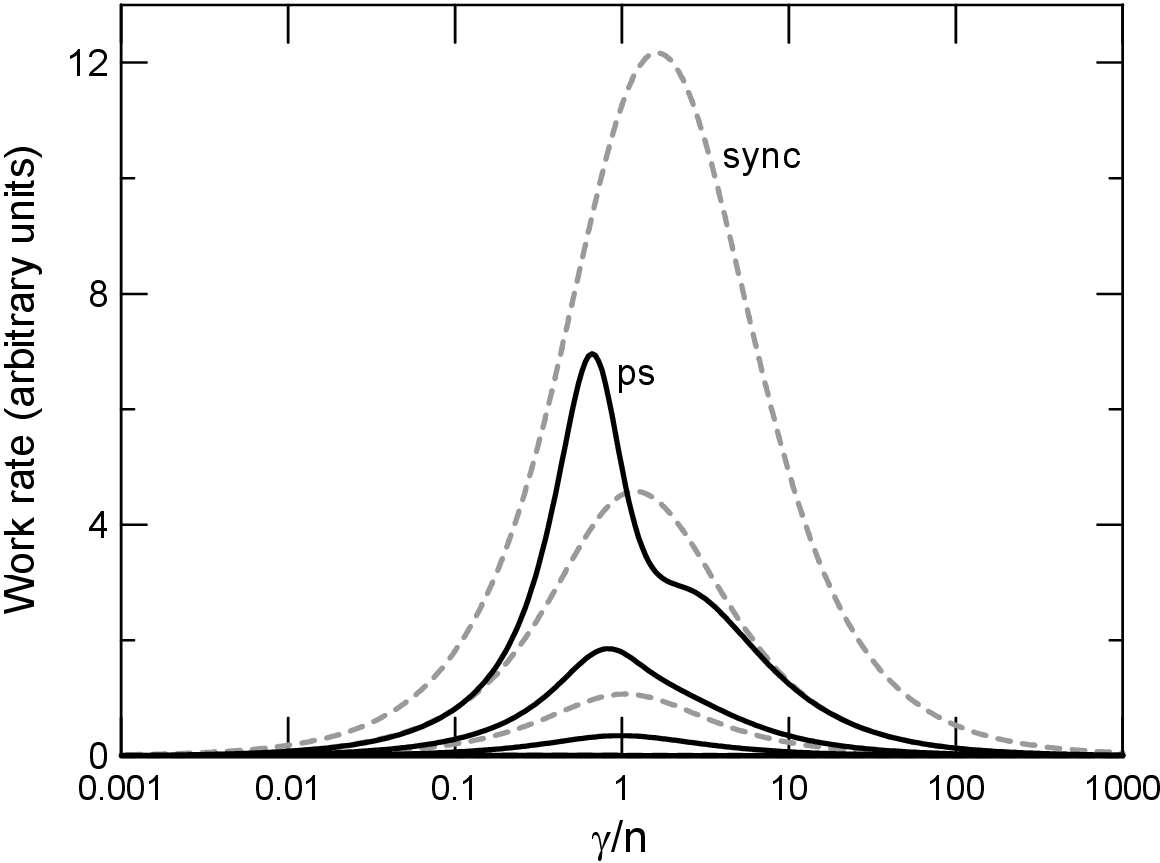}}}
\caption{Time rate of the work done in the synchronous (dashed) and pseudo-synchronous (solid line) solutions for eccentricities between 0.1 (the lowest curves) and 0.3 (the highest curves)} 
\label{fig:Wps}
\end{figure}

\subsection{Free rotating bodies in circular orbits}

In the case of free rotating bodies, the time rate of the work done is given by a series whose leading term is independent of the eccentricity. 
The leading term ($k=0$) is
\beq\label{eq:WdotFree}
\dot{W}_0 \simeq \frac{3GMmn R_e^2\overline\epsilon_\rho}{10a^3} E_{2,0}^2 \sin 2\overline\sigma_0  .
\endeq
A free rotating body in circular motion may show energy variation (at variance with the synchronous and pseudo-synchronous cases in which the energy variation is of order ${\cal O}(e^2)$ and vanish when the orbit is circular).

\subsubsection{Dissipation}

In the case of a body far from a stationary state, however, the energy variation associated with the rotation of $\tens{m}$ may play a major role and needs to be also taken into account in the energy balance. To the variation of the orbital energy, we have to add $\dot{W}_{\rm rot} =C\Omega \dot\Omega$. Hence, using for C the value of the moment of inertia of a homogeneous body, we get the average
\begdi
<\dot{W}_{\rm rot}> = -\ \frac{3GMm\Omega R_e^2\overline\epsilon_\rho }{10a^{3}}\sum_{k\in \Z}
E_{2,k}^2   \sin 2 \overline\sigma_k.
\enddi
and, therefore,
\beq
<\dot{W}_{\rm total}> =-\ \frac{GMm R_e^2\overline\epsilon_\rho }{20a^{3}}\sum_{k\in \Z}
\Big(3(\nu + kn) E_{2,k}^2   \sin 2 \overline\sigma_k + knE_{0,k}^2 \sin 2\overline\sigma''_k  \Big).
\endeq
which is negative (there is a loss of the total mechanical energy) and function of $\nu^2$ (vanishing when $\nu=0$); its modulus is the total energy dissipated inside $\tens{m}$.

It is worth reminding that the behavior of $\dot{\Omega}$ is very complex as discussed in Section \ref{sec:stat}, and that the full equation (Eq. \ref{eq:OdotKep}) may be used if the motion is close to a commensurability where such complexity may affect the result.

\begin{figure}[t]
\centerline{\hbox{
\includegraphics[height=4.5cm,clip=]{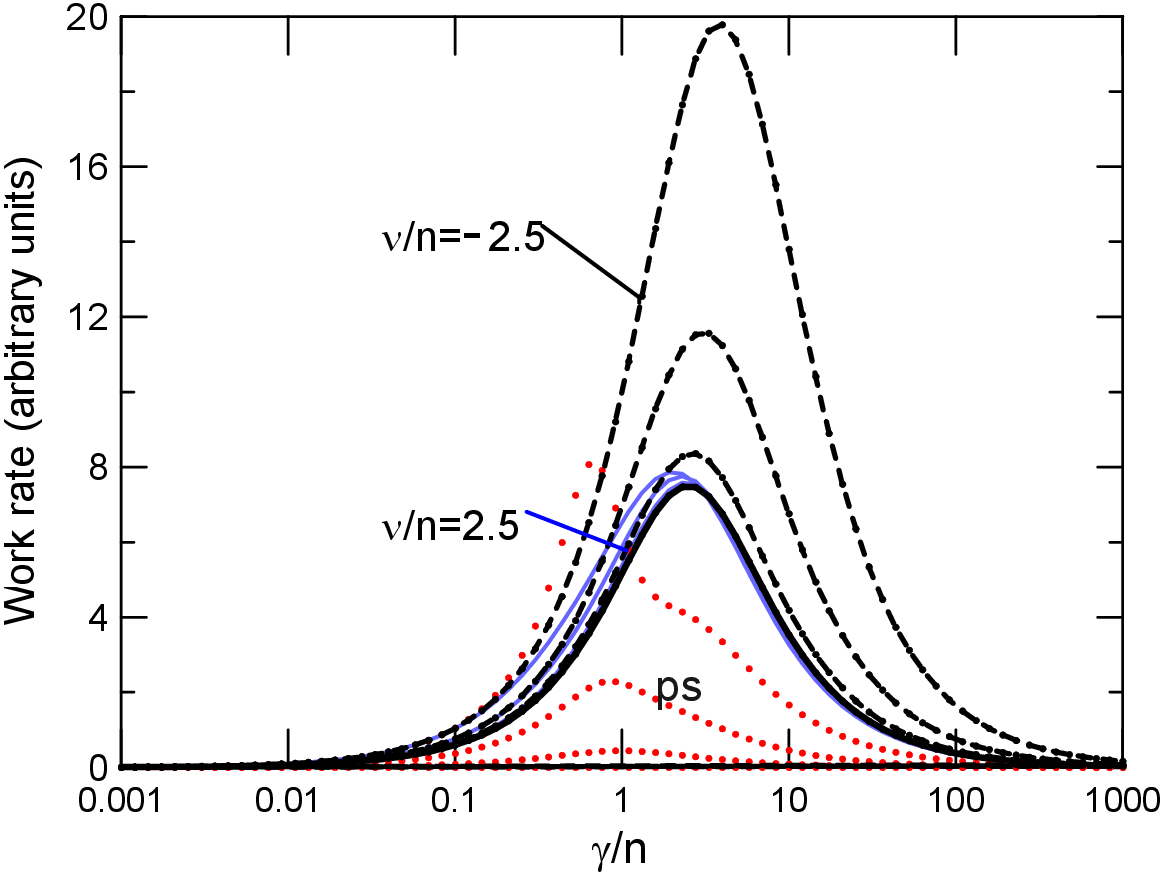}\hspace*{10mm}
\includegraphics[height=4.5cm,clip=]{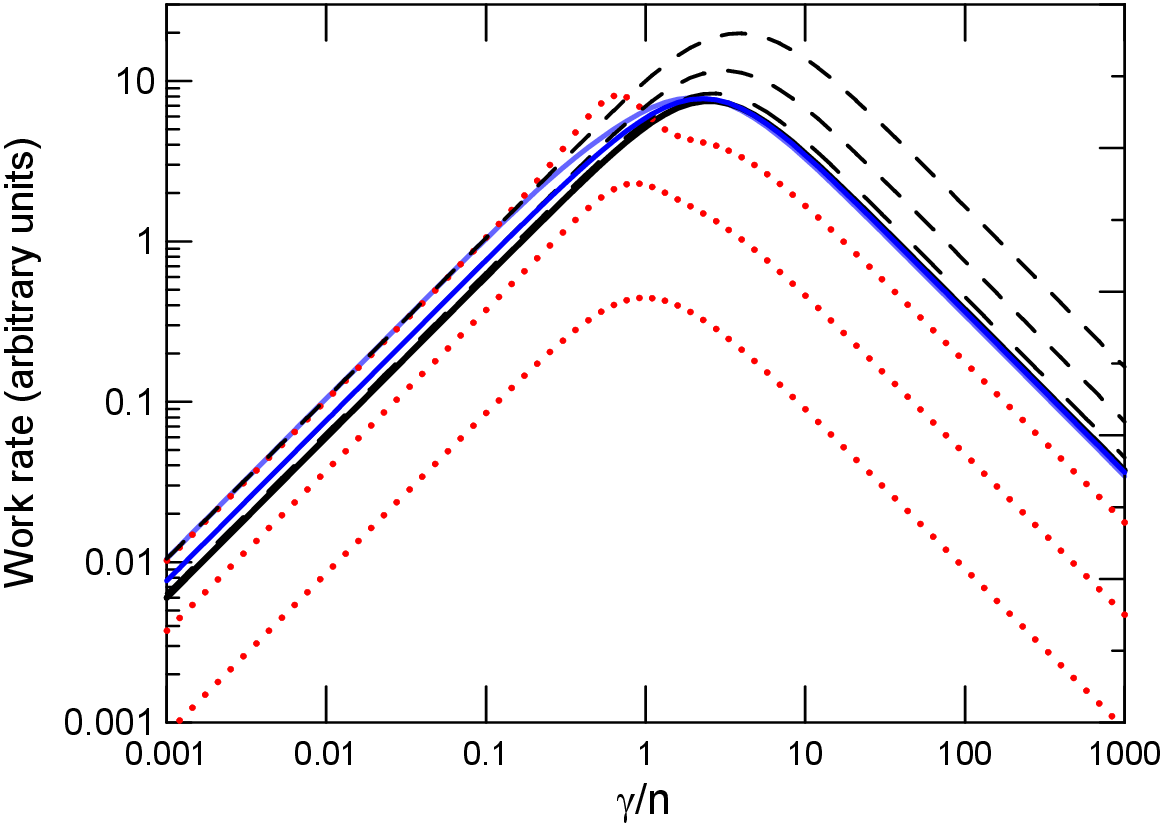}}}
\caption{\textit{Left:}Time rate of the energy dissipated in free rotating bodies in two cases: $\nu=-2.5 n$ (dashed/black) and $\nu=2.5 n$ (solid/blue) for eccentricities between 0.0 (thick line) and 0.3. In the two cases, the results coincide for $e=0$ and the dissipation increases with the eccentricity. The pseudo-synchronous solution is included in the figure (dots/red) for comparison. \textit{Right:} Same in logarithmic scale to show the power laws ruling the dissipation in the two regimes: $\gamma \ll n$ (Efroimsky-Lainey) and $\gamma \gg n$ (Darwin). }
\label{fig:Wfree}
\end{figure}

Figure \ref{fig:Wfree} shows the dissipation in two cases in which $|\nu/n|=2.5$. In the faster case, ($\nu>0$) the body rotation is  {more than twice} the orbital motion; in the other, ($\nu<0$) the rotation is  {slightly retrograde}. The values were chosen so as to avoid being close to the stationary solutions. 

One important inconvenience of the last equation is that it not only is funded on a tide model for homogeneous bodies, but uses also the homogeneity hypothesis in the calculation of the rotational energy of the body. 
This can be partially avoided by using the actual moment of inertia of the body in the calculation of the variation of the rotational energy. If this procedure is used to calculate the energy released in the solid Earth, adopting $\gamma=180 \times 10^{-9}\ {\rm s}^{-1}$, as given in Paper I, and the actual moment of inertia of the Earth, we obtain the release of 150 GW, which is of the same order as the value $110 \pm 25$ GW estimated by Ray et al. (2001) using a Darwinian model and their estimated value for the lag of the semi-diurnal tide. With the same data, Eq. (\ref{eq:WdotFree}) gives 250 GW. Just for completeness, we mention that the value of the solid Earth's 
$\gamma$ given in Paper I is in the border of the confidence interval $240 \pm 60 \times 10^{-9}\ {\rm s}^{-1}$, which we obtain from the value of $Q$ determined by Ray et al. (2001) and the equivalence formulas of Paper I.

\section{Circularization}
Finally, we may consider the variation of the ecccentricity when both the sectorial and zonal components of the creep are considered. The quickest way to get it uses the energy and momentum definitions to obtain, after derivation and elimination of the other variable parameters,
\beq
-\frac{e\dot{e}}{1-e^2}=\frac{\dot{\cal L}}{\cal L}+\frac{\dot{W}}{2W}.
\endeq
After some algebraic manipulatiom, we obtain
\begdi
\dot{e}=-\frac{3GMR_e^2 \overline\epsilon_\rho}{10na^5e} \sum_{k\in \Z}
\Big(2\sqrt{1-e^2}-(2-k-j)(1-e^2)\Big) E_{2,k}\cos\overline\sigma_k 
\sum_{j\in \Z} E_{2,k+j} \sin (j\ell+\overline\sigma_k)
\enddi\beq
-\frac{GMR_e^2 }{10na^5e} \sum_{k\in \Z}
(1-e^2) (k+j) (\overline\epsilon_\rho E_{0,k}+2\delta_{0,k}\overline\epsilon_z)
\cos\overline\sigma''_k 
\sum_{j\in \Z} E_{0,k+j} \sin (j\ell+\overline\sigma''_k)
\endeq
and
\begdi
<\dot{e}>=-\frac{3GMR_e^2 \overline\epsilon_\rho}{20na^5e} \sum_{k\in \Z}
\Big(2\sqrt{1-e^2}-(2-k)(1-e^2)\Big) E_{2,k}^2\sin\ 2\overline\sigma_k 
\enddi\beq
-\frac{GMR_e^2 \overline\epsilon_\rho}{20na^5e} \sum_{k\in \Z}
(1-e^2) k E_{0,k}^2 \sin 2\overline\sigma''_k. 
\endeq

Again, as in the previous sections, the sectorial terms are the same already found in Paper I. The zonal terms are new and, as before, the only contribution comes from the tidal variations of the polar oblateness. The contribution of the term involving the polar oblateness due to the rotation of the body, $\overline\epsilon_z$, vanishes (in the multiplication by $k=0$).

It is important to emphasize that the equations given in this section express the variation of the eccentricity due to the tides raised in $\tens{m}$. The actual variation of the eccentricity may consider also the tides raised in $\tens{M}$, that may be obtained with the same formulas if the meanings of $\tens{m}$ and $\tens{M}$ are interchanged. 

\section{Conclusion}

We have used the creep tide theory to study the evolution of the rotation of bodies due to tides raised in them by the proximity of a second body.
Contrary to the classical spin-orbit studies of solid bodies, no permanent deformation is assumed to exist. The bodies are assumed to behave as viscous fluids free to deform under the tidal action. Any a priori imparted deformation decays exponentially,  by self-gravitation, with a time scale $1/\gamma$. 
The only torques acting on the body result from the asymmetries of its shape generated by the tidal forces. 
It is not said that celestial bodies cannot have a permanent deformation -- the Moon clearly shows such a deformation; they are just not taken into account and we explore to its limits the creep tide of a fluid body. 

The solutions show two distinct patterns. One, when the relaxation factor $\gamma$ is much larger than the frequency of the tide, corresponding to the high-$\gamma$ approximation, dubbed as  Darwin regime, in which the energy dissipation due to the tide is proportional to the tide frequency, and the other, when the relaxation factor is much less than the tide frequency, the Efroimsky-Lainey regime, in which the behavior is inverted and the energy dissipation depends on the frequency of the tide through an inverse power law. Between these two extremes would be the cases in which the relaxation factor and the tide frequency are of the same order, but the known celestial bodies fall allways into one of the two mentioned classes. 
Gaseous planets and stars have relaxation factors of the order of $10 - 100\ {\rm s}^{-1}$ , while terrestrial planets and close-in planetary satellites have relaxation factors of the order of $10 - 100 \times 10^{-9}\ {\rm s}^{-1}$  
We do not know bodies with relaxation factors between these two extremes
On the other hand, the semi-diurnal tide frequencies of the known bodies range from 1.24 $\times 10^{-6}\ {\rm s}^{-1}$  (Mercury) to $180 \times 10^{-6}\ {\rm s}^{-1}$  (Jupiter).  
Thus, we always have either $\gamma \ll \nu$ or $\gamma \gg \nu$. 
The only way to populate the transition zone in which the relaxation factor and the tide frequency are of the same order, would be with very distant synchronous planets or planetary satellites (hypothetical) in which the semi-diurnal frequency would be very small.   

The main topic discussed in the paper is ``synchronization". However, this is a denomination coming from the classical spin-orbit dynamics of solid bodies, which needs to be adapted to be used in the frame of the creep tide theory. 
On one hand, the consiered bodies are not solids, and in the other hand, a damping resulting in a stationary synchronized rotation never happens. 
As shown in the paper, there are many possible behaviors. 
In the simplest Darwinian regime of gaseous bodies, the forced oscillations are damped to negligible values, and the rotation tends indeed to be stationary, but the issue is a bit faster than the orbital motion. The excess of angular velocity is $\sim 6ne^2$ and a synchronous solution only occurs when the eccentricity is zero. 
In the other case, the Efroimsky-Lainey regime of close-in planetary satellites the solutions are damped towards a stationary solution, but it is not necessarily synchronous.
 The stationary solutions found in the paper have periods nearly commensurable with the orbital period, the ratio of the two periods being a half-integer number larger than $-1/2$. If the eccentricity is very small, the only possible solution is indeed nearly synchronous but, as shown in Fig. 2, not exactly synchronized. It is slightly supersynchronous (faster than the mean-motion). 
Besides, this stationary solution only corresponds to the average behavior of the rotation of the body. The actual solution involves important forced oscillations around the attractor.

At last, we have considered the case of some stars in which, besides the tide, we have to consider the possible angular momentum leakage via magnetic stellar wind. In such cases, the star continuous loss of angular momentum displaces the stationary solutions towards sub-synchronous values. Old active host stars (e.g. G stars) with big close-in companions tend to have rotational periods larger than the orbital periods of their companions. On the other hand, non-active host stars (e.g. early F stars) tend to have rotational periods very similar to the orbital periods of their companions. This behavior was verified in a sample of planetary systems with large close-in companions (Ferraz-Mello et al. 2015).
 
The synchronization equation was used to study the rotation of {a molten planet with the same dynamics as} Mercury. It allows the current stationary solution with a period 2/3 of the orbital period (3/2 attractor; $\nu= 0.827 \times 10^{-6}\ {\rm s}^{-1}$) to exist. But it also allows for other possibilities, depending on the relaxation factor of the planet. If we assume that the planet was formed with a rotation much faster than the present and that it was decelerated due to the tide in the planet, it could have been trapped in an attractor corresponding to a faster solution, e,g, with a period of half the orbital period. The fact that it was not trapped in the 2/1 attractor requires that $\gamma > 4.6 \times 10^{-9}\ {\rm s}^{-1}$. In the same way, to remain trapped in the 3/2 attractor we may have $\gamma > 46 \times 10^{-9}\ {\rm s}^{-1}$. 
Besides, the closeness of Mercury's rotation to the exact commensurability indicates a yet stricter limit: $\gamma > 27 \times 10^{-9}\ {\rm s}^{-1}$.

The same study can be extended to some exoplanets. Three examples were discussed in the paper: GJ 667Cc, GJ 581d and Kepler 186 f. The only one showing a possibility of trapping in a non-synchronous attractor is GJ 581d.
However, one necessary condition for this is that it is a super-Earth, but the unknown radius and density of the planet does not allow to exclude the possibility that it is gaseous, in which case $\gamma$ would be high and the rotation would be in Darwin regime where no attractors exist beyond the pseudo-synchronous one.

The paper also includes a discussion of the energy dissipation due to tide raised in the rotating bodies. The general energy variation was discussed in the two main rotation states: pseudo-synchronous and free rotation. If the body is in free rotation, the mechanical energy variation linked to the tide raised on it has two components: the energy variation due to the changes of the system semi-major axis due to the tide, and the variation in the rotational energy of the body. The sum of these two components is the mechanical energy lost by the system and the energy conservation law implies that it may be equal to the thermal energy dissipated inside the body. 
Indeed, similar variations exist in association with the tide raised in the other body, but in the first approximation we may separate the problem in two parts and treat the energy in each body approximating the other body by a material point. In the case of bodies whose rotation is trapped in a stationary state, we can neglect the variation of the rotational energy and consider only the variation due to the changes in the semi-major axis. Actually, the variation in the orbital axis implies a variation in the mean-motion and thus, to remain synchronized, the rotation of the body has also to vary, but the corresponding energy variation is orders of magnitude smaller than the variation of the orbital energy, and can be neglected.

The results showed the Maxwellian pattern already discussed in Paper I. 
In both regimes: $\gamma \gg n$ (Darwin regime) and $\gamma \ll n$ (Efroimsky-Lainey regime), the dissipation follows a power law. See the straight wings in Fig. \ref{fig:Wfree}(\textit {Right}). In the high-$\gamma$, or Darwin regime, the dissipation decreases with $\gamma/n$, or equivalently, grows with the tide frequency ($n/\gamma$). In the language of Darwinian theories, $Q$ is inversely proportional to the frequency or, the geodetic lag is proportional to the frequency). 
For low-$\gamma$, or Efroimsky-Lainey regime, this behavior is inverted. 
The results obtained in the paper were applied to two important examples: Io and the Earth, and were compared with those found in the literature.

In what concerns the version of the creep tide theory used, two points are worth being stressed: (1) We consider, in this version, the effects due to the tidal variation in the polar axis. These variations can be easily understood in the frame of the model used in Paper I, where the polar oblateness of the body was not taken into account. In that case, the equilibrium figure was a prolate Jeans spheroid with the axis directed towards the outer body $\tens{M}$. When the distance to the outer body varies, the prolateness change. This means that the whole section perpendicular to the line joining the two bodies varies; so not only the minor axis in the equatorial section changes, but also the polar axis. So, when we include the polar oblateness due to the rotation of the body, we have also to consider the variation of the polar oblateness due to the tide. (2) We have changed the way in which the forces acting on $\tens{M}$ were computed. This was not a change in the theory since the results obtained here are the same as in Paper I. The interest of the technique used here, in which the deformation of the outer surface is substituted by the variable density of a thin outer layer, opens the possibility of numerical construction of the theory, which may be instrumental when more complex models are considered. 

The model constructed here is yet as simple as possible and just looks for pushing the investigation on the validity limits of the creep tide theory as far as possible before more complex scenarii are considered.  {The elastic component of the tide was not considered because it does not contribute to the torques and energy variation.}

\begin{acknowledgement}
This investigation is funded by the National Research Council, CNPq, grant 306146/2010-0.
\end{acknowledgement}

\section*{Corrections introduced in this version}
\begin{enumerate}
\item Typo in equation (2). The right definition is $\varepsilon_z=1-\frac{c_c}{R_e}$
\item Typo in Eqn. (31). The argument is changed to $k\ell-\overline{\sigma}''_k$ 
\item One radial term is missing in Eqn. (11). The solution including this correction is discussed in the Appendix 1 of Paper III (2017). The changes in the potential are of second order.
\item Mistake in Eqn. (61). In the last line the arguments are changed to  $v+k\ell-\overline{\sigma}''_k$ and $v-k\ell+\overline{\sigma}''_k$
\item Mistake in Eqns. (62-68) The sign in front of the zonal part was wrong. The sign in front of ${\cal C}''_k$ in Eqns. (62-63) is changed to $+$, the sign in front of $kE_{0,k}^2$ in Eqns. (64-66) is changed to $-$ and the sign in in front of $knE_{0,k}^2$ in Eqn. (68) is changed to $+$.
\item Typo in Eqn. (69). The sign in front of the right-hand side is changed to $-$. 
\item Mistakes in Eqns. (70-71) The sign in the beginning of the second lines are changed to $-$.
\item Mistake in Eqn. (B.6) (Online supplement) The sign in front of  
$2\sqrt{1-e^2}E_{2,k}^{(5)}$ is changed to $-$.
\end{enumerate}

\vfill\eject


\vfill\eject

\section*{ONLINE SUPPLEMENT}
\appendix
\renewcommand{\theequation}{A.\arabic{equation}}
\setcounter{equation}{0}  
\renewcommand{\thefigure}{A.\arabic{figure}}
\setcounter{figure}{0}  
  
\section{The equilibrium ellipsoid}\label{ap:Roche}
Let us first remind the equilibrium figure of one body simultaneously submitted to equatorial tide and rotation. 
In the absence of the rotation, the equilibrium figure is approximated by a prolate Jeans spheroid whose major axis is directed to the body $\tens {M}$. 
In the presence of rotation, the equilibrium figure is approximated by a triaxial ellipsoid
whose equatorial prolateness is the same as that of the Jeans spheroid $\epsilon_\rho$ (see Folonier et al. 2015) and whose polar flattening is formed by the flattening of the Maclaurin spheroid plus the contribution of the tidal potential (which depends on the distance $r$). 
With respect to the principal axes, the equation of the adopted triaxial ellipsoid is 
\beq
\frac{X^2}{a^2} + \frac{Y^2}{b^2} +  \frac{Z^2}{c^2} = 1
\endeq
where $a=R_e(1+\half \epsilon_\rho), b=R_e(1-\half \epsilon_\rho), c=R_e(1-\epsilon_z)$ (N.B. $a>b>c$),  $R_e$ is the mean equatorial radius. 
One  rotation is needed to bring the system whose axes are the principal axes into the adopted system of reference. 
The coordinates of one point on the surface of the body, in the reference system fixed in the body, are 
 $x=\rho\sin\widehat\theta\cos\widehat\varphi_F,  y=\rho\sin\widehat\theta\sin\widehat\varphi_F, z=\rho\cos\widehat\theta$
where $\widehat\theta$ and $\widehat\varphi_F$ are, respectively, the co-latitude and longitude of the point.
The sought rotation shifts the origin of the system of coordinates to the axis $a$ (i.e. towards $\tens{M}$). 
The angle between the two systems is $\varpi+v$ where $\varpi$ is the longitude of the pericenter counted from the origin fixed in the rotating body and $v$ is the true anomaly of the body $\tens{M}$.
We thus have
\beq\label{eq:roche}
\left(\begin{array}{c}X\\Y\\Z\end{array}\right)=
\left(\begin{array}{c@{\quad}c@{\quad}c}\cos(\varpi+v)&\sin(\varpi+v)&0\\ 
-\sin(\varpi+v)&\cos(\varpi+v)&0\\0&0&1\end{array}\right)
\left(\begin{array}{c}x\\y\\z\end{array}\right),
\endeq
or
\beq\begin{array}{l@{\speq}l@{}l}
X & \rho \sin\widehat\theta \cos(\widehat\varphi_F - \varpi-v) & \speq \rho \sin\widehat\theta \cos(\widehat\varphi - \omega-v)\\
Y & \rho \sin\widehat\theta \sin(\widehat\varphi_F - \varpi-v) & \speq  \rho \sin\widehat\theta \sin(\widehat\varphi - \omega-v)\\
Z & \rho  \cos\widehat\theta \\
\end{array}\endeq
where $\widehat\varphi$ and $\omega$ are counted from a fixed direction in space (corresponding to the nodal line in the spatial theory) (see Fig. 1).

Hence, after substitution in the equation of the ellipsoid,
\beq
\rho=R_e \Big(1 + \half \epsilon_\rho \sin^2\widehat\theta \cos (2\widehat\varphi - 2\omega - 2v) - \epsilon_z \cos^2\widehat\theta  \Big)
\endeq

The polar oblateness is composed of a part fixed by the rotation of the body (Maclaurin) and the increment related to the tidal deformation:
\beq
{\epsilon_z}=\overline\epsilon_z+\half\epsilon_\rho
\endeq
In absence of rotation, $\epsilon_z=\half\epsilon_\rho$ (Jeans prolate spheroid). 

\renewcommand{\theequation}{B.\arabic{equation}}
\setcounter{equation}{0}  
\section{Cayley expansions}\label{sec:Cayley}

\subsection{Complete Cayley functions}
In the paper, we only use the Cayley functions defined by the third power of $a/r$. However, the functions defined by Cayley (1861) are more general:
\begdi
E^{(n)}_{q,p}(e)=\frac{1}{2\pi}\int_0^{2\pi}\left(\frac{a}{r}\right)^n
\cos\big(qv+(p-q)\ell\big)\ d\ell.
\enddi

Cayley functions defined by the higher powers of $a/r$ can be calculated from those defined by the  lower powers through the recurrence formula
\begdi 
E^{(n+1)}_{q,p} = \frac{1}{1-e^2} \Big(E^{(n)}_{q,p}+\frac{e}{2}(E^{(n)}_{q+1,p+1}+E^{(n)}_{q-1,p-1})\Big).
\enddi


\subsection{Basic formulas}
\subsubsection{Case I}
Given the function
\beq
F=\frac{a^3}{r^3}\sin(qv-p\ell+\Phi) \hspace{3cm} q\in\mathbb{Z},
\endeq
its Fourier expansion is
\begdi
F=\frac{1}{2\pi}\int_{-\pi}^\pi F d\ell + \sum_{k=1}^\infty \Big(
\frac{1}{\pi} \cos k\ell \int_{-\pi}^\pi F \cos k\ell\ d\ell+
\frac{1}{\pi} \sin k\ell \int_{-\pi}^\pi F \sin k\ell\ d\ell \Big)
\enddi
where the trigonometric functions of every term may be decomposed into their parts dependent and independent on $\ell$ (including $v(\ell)$). 
Keeping only the even terms\footnote{The integral from $-\pi$ to $+\pi$ of the odd terms is equal to zero}, we obtain:
\begdi
F=\frac{1}{\pi}  \sin\Phi\int_{-\pi}^\pi \frac{a^3}{r^3} \cos(qv-p\ell) d\ell +
 \sum_{k=1}^\infty\Big(
\frac{1}{\pi} \cos k\ell  \sin\Phi\int_{-\pi}^\pi \frac{a^3}{r^3} \cos(qv-p\ell)\cos k\ell 
\ d\ell +
\enddi\begdi
\frac{1}{\pi} \sin k\ell \cos\Phi\int_{-\pi}^\pi \frac{a^3}{r^3} \sin(qv-p\ell) \sin k\ell 
\ d\ell \Big) \enddi
or
\begdi
F=\frac{1}{2\pi}  \sin\Phi\int_{-\pi}^\pi \frac{a^3}{r^3} \cos(qv-p\ell) \ d\ell
 + \sum_{k=1}^\infty \Big(
\frac{1}{2\pi} \cos k\ell  \sin\Phi\int_{-\pi}^\pi \frac{a^3}{r^3} 
\big(\cos(qv-p\ell +k\ell) +\cos(qv-p\ell - k\ell) \big)\ d\ell +  \enddi\begdi
\frac{1}{2\pi} \sin k\ell \cos\Phi\int_{-\pi}^\pi \frac{a^3}{r^3} 
\big(\cos(qv-p\ell -k\ell) - \cos(qv-p\ell + k\ell)\big) \ d\ell\Big)
\enddi
The resulting integrals are the Cayley coefficients, after the introduction of which the function $F$ becomes
\begdi
F=\sin\Phi E_{q,q-p}      +   \sum_{k=1}^\infty \Big(
\cos k\ell  \sin\Phi \big(E_{q,q-p+k}+E_{q,q-p-k}\big) +
\sin k\ell  \cos\Phi \big(E_{q,q-p-k}-E_{q,q-p+k}\big) \Big)
\enddi
or
\begdi
F= \sum_{k=-\infty}^\infty E_{q,q-p+k} \sin(\Phi-k\ell)
\enddi
and
\begdi
<F>= E_{q,q-p} \sin\Phi
\enddi
\subsubsection{Case II}
Consider the function
\beq
F'=\frac{a^3}{r^3}\cos(qv-p\ell+\Phi) \hspace{3cm} q\in\mathbb{Z},
\endeq
Hence
\begdi
F'=\frac{a^3}{r^3}\sin(qv-p\ell+\Phi+\frac{\pi}2)
\enddi
and from the above result:
\begdi
F'= \sum_{k=-\infty}^\infty E_{q,q-p+k} \sin(\Phi-k\ell+\frac{\pi}2)
\enddi
or
\beq
F'= \sum_{k=-\infty}^\infty E_{q,q-p+k} \cos(\Phi-k\ell)
\endeq
and
\beq
<F'>= E_{q,q-p} \cos\Phi
\endeq

\subsection{Auxiliary formulas}
We consider in this appendix two formulas used to express the energy dissipation in terms of the usual Cayley functions (i.e., those for the third power of $a/r$).
\subsubsection*{Proposition B.3.1}
\beq
kE^{(3)}_{0,k}= \frac{3e}{2\sqrt{1-e^2}} \left(E^{(4)}_{1,1-k} - E^{(4)}_{1,1+k}\right)
\endeq

From the definition of the Cayley functions, we may write
\begdi
kE^{(3)}_{0,k}=\frac{1}{2\pi}\int_{\ell=0}^{\ell=2\pi}\left(\frac{a}{r}\right)^3
d (\sin k\ell) = \Big[\left(\frac{a}{r}\right)^3 \sin k\ell \Big]_0^{2\pi}
-\frac{1}{2\pi}\int_{\ell=0}^{\ell=2\pi}  \sin k\ell\  d\left(\frac{a}{r}\right)^3.
\enddi

Using the differential form of the area's law, we obtain,
\begdi
d\left(\frac{a}{r}\right)^3=-3 \left(\frac{a}{r}\right)^2 \frac{e}{1-e^2} \sin v\ dv = 
-3 \left(\frac{a}{r}\right)^4 \frac{e} {\sqrt{1-e^2}} \sin v\ d\ell
\enddi
and so,
\begdi
kE^{(3)}_{0,k}=\frac{3e}{2\sqrt{1-e^2}} \frac{1}{2\pi}\int_{0}^{2\pi}\left(\frac{a}{r}\right)^4 
\big(\cos (v-k\ell) - \cos (v+k\ell)\big) d\ell. 
\enddi
\flushright{\qed} 

\flushleft
\subsubsection*{Proposition B.3.2}
\beq
(k-2)E^{(3)}_{2,k}=-\frac{3e}{2\sqrt{1-e^2}} (E^{(4)}_{3,k+1} -  E^{(4)}_{1,k-1}) 
-2\sqrt{1-e^2}  E^{(5)}_{2,k} 
\endeq

We proceed as in the previous Proposition; the only difference is that we have to decompose the trigonometric argument; hence 
\begdi
(k-2)E^{(3)}_{2,k}
=\frac{1}{2\pi}\int_{\ell=0}^{\ell=2\pi}\left(\frac{a}{r}\right)^3 
\big(\cos 2v\ d[\sin (k-2)\ell] + \sin 2v\ d[\cos (k-2)\ell]\big) 
\enddi
\begdi
=-\frac{1}{2\pi}\int_{\ell=0}^{\ell=2\pi}
\left(\sin (k-2)\ell\ d\left[\left(\frac{a}{r}\right)^3 \cos 2v \right] 
+\cos (k-2)\ell\ d\left[\left(\frac{a}{r}\right)^3 \sin 2v \right] \right).
\enddi
The differentials in this case are:
\begdi
d\left[\left(\frac{a}{r}\right)^3\cos 2v\right]= -3\left(\frac{a}{r}\right)^4 \frac{e}{\sqrt{1-e^2}} \sin v \cos 2v\ d\ell - 2 \sqrt{1-e^2}\left(\frac{a}{r}\right)^5 \sin 2v\ d\ell
\enddi
\begdi
d\left[\left(\frac{a}{r}\right)^3\sin 2v\right]= -3\left(\frac{a}{r}\right)^4 \frac{e}{\sqrt{1-e^2}} \sin v \sin 2v\ d\ell + 2 \sqrt{1-e^2}\left(\frac{a}{r}\right)^5 \cos 2v\ d\ell;
\enddi
substituting and regrouping the terms, we obtain
\begdi
(k-2)E^{(3)}_{2,k}=-\frac{1}{2\pi}\int_{0}^{2\pi}\Big(
\frac{3e}{2\sqrt{1-e^2}} \left(\frac{a}{r}\right)^4 
\big(\cos (3v+(k-2)\ell) - \cos (v+(k-2)\ell)\big) 
\enddi
\begdi
+2\sqrt{1-e^2} \left(\frac{a}{r}\right)^5 
\cos (2v+(k-2)\ell)\Big)d\ell. 
\enddi
\flushright{\qed} 
\flushleft

\subsection{Cayley coefficients}

\begin{table}[h!]
\label{tab:Ek}
\caption{Cayley coefficients $E_{2,k}$. }
\beq\begin{array}{l@{\speq}l}
E_{2,-7} & \displaystyle \frac{12144273}{71680} e^7\nonumber\vspace*{1mm} \\
E_{2,-6} & \displaystyle \frac{73369}{720} e^6 \nonumber\vspace*{1mm} \\
E_{2,-5} & \displaystyle \frac{228347}{3840} e^5 - \frac{3071075}{18432} e^7\nonumber\vspace*{1mm} \\
E_{2,-4} & \displaystyle \frac{533}{16} e^4 - \frac{13827}{160} e^6 \nonumber\vspace*{1mm} \\
E_{2,-3} & \displaystyle \frac{845}{48} e^3 - \frac{32525}{768} e^5 + \frac{208225}{6144} e^7\nonumber\vspace*{1mm} \\
E_{2,-2} & \displaystyle \frac{17}{2} e^2 - \frac{115}{6} e^4 + \frac{601}{48} e^6 \nonumber\vspace*{1mm} \\
E_{2,-1} & \displaystyle \frac{7}{2} e - \frac{123}{16} e^3 + \frac{489}{128} e^5 - \frac{1763}{2048} e^7\nonumber \vspace*{1mm} \\
E_{2,0} & \displaystyle 1 - \frac{5}{2} e^2 + \frac{13}{16} e^4 - \frac{35}{288} e^6\nonumber \vspace*{1mm} \\
E_{2,1} & \displaystyle -\frac{1}{2} e + \frac{1}{16} e^3 - \frac{5}{384} e^5 - \frac{143  }{18432} e^7\nonumber\vspace*{1mm} \\
E_{2,2} & \displaystyle 0\nonumber\vspace*{1mm} \\
E_{2,3} & \displaystyle \frac{1}{48} e^3 + \frac{11}{768} e^5 + \frac{313}{30720} e^7\nonumber\vspace*{1mm} \\
E_{2,4} & \displaystyle \frac{1}{24} e^4 + \frac{7}{240} e^6\nonumber\vspace*{1mm} \\
E_{2,5} & \displaystyle \frac{81}{1280} e^5 + \frac{81}{2048} e^7\nonumber\vspace*{1mm} \\
E_{2,6} & \displaystyle \frac{4}{45} e^6 \nonumber\vspace*{1mm}\\
E_{2,7} & \displaystyle \frac{15625}{129024} e^7\nonumber \\
\end{array}
\endeq
\end{table}

\begin{table}[h!]
\label{tab:E0k}
\caption{Cayley coefficients $E_{0,k}$. }
\beq\begin{array}{l@{\speq}l}
E_{0,0 } & \displaystyle 1 + \frac{3}{2} e^2 + \frac{15}{8} e^4 - \frac{35}{16} e^6\nonumber \vspace*{1mm} \\
E_{0,1 } & \displaystyle \frac{3}{2} e + \frac{27}{16} e^3 - \frac{261}{128} e^5 + \frac{14309}{6144}e^7\nonumber\vspace*{1mm} \\
E_{0,2 } & \displaystyle   \frac{9}{4}e^2+\frac{7}{4}e^4+\frac{141}{64}e^6
\nonumber\vspace*{1mm} \\
E_{0,3 } & \displaystyle \frac{53}{16} e^3 + \frac{393}{256} e^5 + \frac{24753}{10240} e^7\nonumber \vspace*{1mm} \\
E_{0,4 } & \displaystyle \frac{77}{16} e^4 + \frac{129}{160} e^6\nonumber\vspace*{1mm} \\
E_{0,5 } & \displaystyle \frac{1773}{256} e^5 - \frac{4987}{6144} e^7\nonumber\vspace*{1mm} \\
E_{0,6 } & \displaystyle \frac{3167}{320} e^6 \nonumber\vspace*{1mm}\\
E_{0,7 } & \displaystyle \frac{432091}{30720} e^7\nonumber \\
\end{array}
\endeq
N.B. E$_{0,-k}=E_{0,k}$
\end{table}

\begin{table}[t!]
\label{tab:E4k}
\caption{Cayley coefficients $E_{4,k}$.} 
\beq\begin{array}{l@{\speq}l}

E_{4,-7}  & \displaystyle \frac{131087143}{129024} e^7\nonumber\vspace*{1mm} \\
E_{4,-6}  & \displaystyle \frac{75947}{144} e^6 \nonumber\vspace*{1mm} \\
E_{4,-5}  & \displaystyle \frac{333513}{1280} e^5 - \frac{18298713}{10240} e^7\nonumber\vspace*{1mm} \\
E_{4,-4}  & \displaystyle \frac{2893}{24} e^4 - \frac{40387}{48} e^6 \nonumber\vspace*{1mm} \\
E_{4,-3}  & \displaystyle \frac{2443}{48} e^3 - \frac{284557}{768} e^5 + \frac{29629663}{30720} e^7\nonumber\vspace*{1mm} \\
E_{4,-2}  & \displaystyle \frac{75}{4} e^2 - \frac{595}{4} e^4 + \frac{12513}{32} e^6 \nonumber\vspace*{1mm} \\
E_{4,-1}  & \displaystyle \frac{11}{2} e - \frac{833}{16} e^3 + \frac{55135}{384} e^5 - \frac{2975165}{18432} e^7\nonumber \vspace*{1mm} \\
E_{4,0 } & \displaystyle 1 - \frac{29}{2} e^2 + \frac{365}{8} e^4 - \frac{7111}{144} e^6\nonumber \vspace*{1mm} \\
E_{4,1 } & \displaystyle -\frac{5}{2} e + \frac{183}{16} e^3 - \frac{1611}{128} e^5 - \frac{8035}{2048}e^7\nonumber\vspace*{1mm} \\
E_{4,2 } & \displaystyle \frac{7}{4}e^2-\frac{29}{12}e^4+\frac{1}{3}e^6\nonumber\vspace*{1mm} \\
E_{4,3 } & \displaystyle -\frac{13}{48} e^3 - \frac{29}{768} e^5 - \frac{197}{6144} e^7\nonumber\vspace*{1mm} \\
E_{4,4 } & \displaystyle 0 \nonumber\vspace*{1mm} \\
E_{4,5 } & \displaystyle \frac{11}{3840} e^5 + \frac{407}{9216} e^7\nonumber\vspace*{1mm} \\
E_{4,6 } & \displaystyle \frac{1}{288} e^6 \nonumber\vspace*{1mm}\\
E_{4,7 } & \displaystyle \frac{243}{71680} e^7\nonumber \\
\end{array}
\endeq
\end{table}

\end{document}